\documentclass{article}

\usepackage{arxiv}

\usepackage[utf8]{inputenc}
\usepackage[T1]{fontenc}
\usepackage[english]{babel}
\usepackage{graphicx}	
\usepackage{amsmath}	
\usepackage{amssymb}	
\usepackage{color}
\usepackage{fullpage}
\usepackage{morefloats}
\usepackage{lscape}
\usepackage{fancyhdr}
\usepackage{rotating}
\usepackage{bookmark}
\usepackage{latexsym}
\usepackage{xspace}
\usepackage{enumitem}
\usepackage{mathtools}
\usepackage{xparse}
\usepackage{array, booktabs}

\usepackage{natbib}

\newcommand\aap{A\&A}                
\newcommand\afz{Afz}                 
\newcommand\apj{ApJ}                 
\newcommand\azh{Azh}                 
\newcommand\mnras{MNRAS}             
\newcommand\prb{Phys. Rev.~B}        

\newcommand\nifsx{${}^{56}$Ni\xspace}
\newcommand\cofsx{${}^{56}$Co\xspace}
\newcommand\fefsx{${}^{56}$Fe\xspace}
\newcommand\chainnicofe{\mbox{\nifsx $\Rightarrow$ \cofsx $\Rightarrow$ \fefsx}\xspace}
\newcommand{\code}[1]{\texttt{#1}}
\newcommand{\WI}[2]{#1_{\mathrm{#2}}}

\ExplSyntaxOn

\makeatletter
\NewDocumentCommand{\multicitep}{m}
 {
  \NAT@open
  \mjb_multicitep:n { #1 }
  \NAT@close
 }
\makeatother

\seq_new:N \l_mjb_multicite_in_seq
\seq_new:N \l_mjb_multicite_out_seq
\seq_new:N \l_mjb_cite_seq

\cs_new_protected:Npn \mjb_multicitep:n #1
 {
  \seq_set_split:Nnn \l_mjb_multicite_in_seq { ; } { #1 }
  \seq_clear:N \l_mjb_multicite_out_seq
  \seq_map_inline:Nn \l_mjb_multicite_in_seq
   {
    \mjb_cite_process:n { ##1 }
   }
  \seq_use:Nn \l_mjb_multicite_out_seq { ;~ }
 }

\cs_new_protected:Npn \mjb_cite_process:n #1
 {
  \seq_set_split:Nnn \l_mjb_cite_seq { , } { #1 }
  \int_compare:nTF { \seq_count:N \l_mjb_cite_seq == 1 }
   {
    \seq_put_right:Nn \l_mjb_multicite_out_seq
     { \citeauthor{#1},~\citeyear{#1} }
   }
   {
    \seq_put_right:Nx \l_mjb_multicite_out_seq
     {
      \exp_not:N \citeauthor{\seq_item:Nn \l_mjb_cite_seq { 1 }},~
      \exp_not:N \citeyear{\seq_item:Nn \l_mjb_cite_seq { 1 }},~
      \seq_item:Nn \l_mjb_cite_seq { 2 }
     }
   }
 }
\ExplSyntaxOff

\title{A simple model of time-dependent ionization in~Type~IIP supernova envelope}

\date{2019 year}

\author
{
  M.Sh.~Potashov \\
  NRC ``Kurchatov Institute'' - ITEP, ul. Bolshaya Cheremushkinskaya 25, Moscow 117218, Russia\\
  Novosibirsk State University, Pirogova 1, Novosibirsk 630090, Russia\\
  \texttt{marat.potashov@gmail.com} \\
  \And
  A.V.~Yudin \\
  NRC ``Kurchatov Institute'' - ITEP, ul. Bolshaya Cheremushkinskaya 25, Moscow 117218, Russia\\
  NRC ``Kurchatov Institute'', Akademika Kurchatova pl. 1, Moscow 123182, Russia\\
  \texttt{yudin@itep.ru} \\
}

\begin{document}

\maketitle

\begin{abstract}
  We propose a model kinetic system of the hydrogen atom
    (two levels plus continuum) under the conditions typical
    for atmospheres of Type IIP supernovae in the plateau stage.
  Despite the simplicity of this system,
    it describes realistically the basic properties
    of the complete system.
  Analysis shows that the ionization ``freeze-out'' effect
    is always manifest at large times.
  We give a simple criterion for checking the statistical
    equilibrium of a system under the given conditions at any time.
  It is shown that if the system is in non-equilibrium at early times,
    the time-dependent effect of ionization necessarily exists.
  We also generalize this criterion to the case of
    arbitrary multilevel systems.
  We discuss various factors that affect the strength of the
    time-dependent effect in the kinetics during
    the photospheric phase of a supernova explosion.
\end{abstract}


\section{Introduction}
  \label{sec:introduction}

The investigation of the structure of the Universe
  involves the measurement of photometric distances to
  objects with known redshifts.
There are a large variety of methods for measuring distances.
Among them, there are methods that do not rely on
  the cosmological distance ladder,
  such as the expanding photosphere method (EPM)
  \citep{KirshnerKwan1974},
  the spectral-fitting expanding atmosphere method (SEAM)
  \citep{BaronNugentBranchEtal2004},
  and the dense shell dethod (DSM)
  \citep{BlinnikovPotashovBaklanovEtal2012,
    PotashovBlinnikovBaklanovEtal2013,
    BaklanovBlinnikovPotashovEtal2013a}.
These methods are based on the properties of Type~IIP and Type~IIn supernovae
  (SNe~IIP and SNe~IIn).
Some of them (e.g., SEAM) are very complicated and require full physical
  modelling of the SN with a detailed reproduction of its spectrum.
Direct cosmological methods for distance measurement
  are especially important owing to the problem of the uncertainty
  in the Hubble parameter (Hubble tension)
  \citep{RiessCasertanoYuanEtal2018,
    MortsellDhawan2018,
    EzquiagaZumalacarregui2018}.

In order to model the physical processes occurring within SN ejecta,
  it is necessary to solve a system of partial integro-differential equations
  of radiation hydrodynamics, which includes
  the envelope expansion hydrodynamics,
  the interaction of the radiation field with matter,
  the radiation transfer in lines and in the continuum,
  and the kinetics of level populations in atoms of multiply charged plasma.
The complete numerical solution of this system
  is still an impossible task even in the one-dimensional case,
  and one has to resort to unavoidable simplifications.
An important and frequently used simplification is the steady-state approximation
  for the kinetic system of level populations, when the system is assumed
  to be in statistical equilibrium.
The effect of time-dependence is manifest
  in the deviation of the actual occupation numbers
  of the atomic levels from their steady-state (equilibrium) values.

The time-dependent effect of hydrogen ionization in the
  envelopes of SNe~II in the photospheric phase
  was used by \citet{KirshnerKwan1975}
  to explain the high H$\alpha$ intensity in the spectra of SN~1970G,
  and by
  \citet{Chugai1991}
  to explain the high degree of hydrogen excitation
  in the outer atmospheric layers ($v>7000$ km\,s$^{-1}$)
  of SN~1987A for the first 40 d after the explosion.

\citet{UtrobinChugai2002}
  found a strong time-dependent effect in the ionization kinetics
  and hydrogen lines in SNe~IIP during the photospheric phase.
In the next paper,
  \citet{UtrobinChugai2005}
  also took into account the time-dependent effect in the energy equation.
In these papers, it was shown that
  the H$\alpha$ line was enhanced in the spectrum of the peculiar
  SN~1987A owing to the time-dependent ionization.
In the steady-state approximation, this effect has been achieved only by mixing
  radioactive \nifsx into the outer high-velocity layers.
\citet{Utrobin2007} showed
  the importance
  of non-stationarity also for a normal SN~IIP, SN~1999em.

The conclusion of Utrobin and Chugai
  was confirmed by Dessart and Hillier using the \code{CMFGEN} software package.
\citet{DessartBlondinBrownEtal2008} still applied the steady-state approach
  that was implemented in the \code{CMFGEN} package,
  but the H$\alpha$ line in the hydrogen-rich envelopes
  was weaker than the observed one during the recombination epoch.
In particular, the model did not reproduce H$\alpha$ after the SN age of
  4~d for SN~1987A and 20~d for SN~1999em.
Another version of \code{CMFGEN} was improved by including the time dependence
  in the kinetic system and the energy equation
  \citep{DessartHillier2007},
  and for the latest version, that in the radiative transfer
  \citep{DessartHillier2010a, HillierDessart2012}.
This strengthened the H$\alpha$ in the
  simulated spectrum and led to better agreement with observations.

On the other hand, based on the computations
  with the \code{PHOENIX} software package,
  \citet{DeBaronHauschildt2010a}
  found that the time-dependent kinetics
  is important only during the first few days after the SN explosion.
Moreover, they argued that the role of the time-dependent effect
  is not very strong even in the first few days,
  and illustrated this with models
  of SN~1987A and SN~1999em.

\citet{VoglSimNoebauerEtal2019},
  recognized the importance of time-dependent effect
  for the ionization kinetics, but neglected it.
Nevertheless, they obtained a good agreement
  of the spectra of the SN~1999em simulated
  with the open source code \code{TARDIS} with the observed ones.
The vast majority of Monte Carlo simulation codes also neglect the
  time-dependent effect in kinetics.
Thus, the conclusions of the various research groups disagree,
  and the importance of the effect is still being debated.

\citet{PotashovBlinnikovUtrobin2017a}
  showed the importance of the non-stationary kinetics
  for SN~1999em in the purely hydrogen case
  using the codes \code{STELLA}
  \citep{BlinnikovEastmanBartunovEtal1998,
    BlinnikovLundqvistBartunovEtal2000,
    BlinnikovRopkeSorokinaEtal2006}
  and \code{LEVELS}.
The influence of metal admixtures on the non-stationarity was also studied.
The increase of the metal abundance in the envelope
  led to a weakening of the time-dependent effect.

In this paper, we use a simple analytical model to determine
  whether and when the time-dependent ionization effect is important.
We give a simple criterion for checking the statistical equilibrium of a system.
We distinguish two important periods
  in the evolution of the system under study.
The first period starts from the beginning of the
  photospheric phase
  and lasts not longer than the time-scale
  for the change of the photospheric parameters,
We refer to this period as ``early time''.
Typically, it lasts for a few days.
Much longer periods will be called ``large time''.

A brief outline of the paper is as follows.
Section~\ref{sec:model} describes the
  physical model of the problem.
In Section~\ref{sec:large_time}, it is concluded
  that the time-dependent effect is significant at least
  in the large time limit.
In Section~\ref{sec:early_time}, the evolution of
  deviation from stationarity
  with time is studied, assuming that initially it is small.
A formula for this evolution is derived.
This formula depends on an expression that defines the strength
  of the time-dependent effect.
In Section~\ref{sec:frozen_system},
  it is shown that the expression described in Section~\ref{sec:early_time} gives
  a simple criterion for checking the statistical equilibrium of the system
  under the given conditions at any time.
For the system in equilibrium, there will be no time-dependent effect.
In this section, we also conclude
  that the key factor affecting the relaxation time
  of the entire system is the intensity
  and rate of change of the external hard continuum radiation
  between the Lyman and Balmer ionization thresholds.
Finally, in Section~\ref{sec:full_system} we investigate
  various kinetic systems using a generalized equilibrium criterion
  and find additional factors that influence
  the effect of non-stationary ionization.
We also discuss how this effect is affected
  by the metallicity of the stellar envelope.

\section{Physical model}
  \label{sec:model}

We consider a reasonably simple analytical model
  for the behaviour of level populations
  of a purely hydrogen plasma
  in the supernova envelope.
  The hydrogen atom is represented by the system
    ``two levels plus continuum''.
We assume an $l$-equilibrium in the kinetic system for the second level.
This means that the populations of the fine-structure sublevels
  $2\mathrm{s}$ and $2\mathrm{p}$
  are proportional to their statistical weights.
Thus, the second level is considered as a single so-called super-level
  \citep{HubenyLanz1995}.

The light-curve behaviour of a typical SN~IIP
  can be divided into various characteristic stages
  \citep{Utrobin2007}:
\begin{itemize}
  \item shock breakout;
  \item adiabatic cooling phase;
  \item photospheric phase (cooling and recombination\\ wave);
  \item phase of radiative diffusion cooling;
  \item exhaustion of radiation energy;
  \item plateau tail;
  \item radioactive tail (\chainnicofe).
\end{itemize}

To study the time-dependent ionization of the hydrogen plasma,
  we will consider the behaviour of the system only in the photospheric phase.
For the typical supernova SN~1999em
  \citep{BaklanovBlinnikovPavlyuk2005, Utrobin2007}
  this phase lasts from ${t_0 \sim 20}$~d to ${\sim 100}$~d.
As the envelope expands, a cooling and recombination wave is formed.
The bolometric luminosity of the SN
  is equal to the luminosity at the outer edge of this wave.
The photosphere is located at the same level.
It is important that during the photospheric phase, the
  photospheric radius $\WI{R}{ph}$,
  the radiation temperature $T_c$
  and the gas temperature $\WI{T}{e}$
  are nearly constant.
Consequently, the luminosity of the SN does not change in time,
  and one observes a plateau in the light-curve.

A radiation-hydrodynamical simulation of the SN~1999em envelope
  with the code \code{STELLA} shows that the transition
  to homologous expansion (with high accuracy)
  is completed by about 15~d after the explosion
  \citep{BaklanovBlinnikovPavlyuk2005},
  which is before the beginning of the photospheric phase $t_0$.
We assume that the gas expands isotropically
  (i.e. a one-dimensional spherically symmetric approximation is used)
  and do not take collisional processes into account.
The validity of the latter assumption is discussed in Section~\ref{sec:full_system}
  (see also Appendix~\ref{ap:tube}).

Let us select a small area of the envelope above the photosphere.
The continuity equation in Eulerian coordinates for the gas in this region is
\begin{equation}\label{eq:continuity:eulerian}
  \frac{\partial\rho}{\partial t} = -\bigtriangledown(\rho v),
\end{equation}
  where $\rho$ is the density of the envelope expanding with a velocity $v$.
In the Lagrangian formalism in the comoving frame, we obtain
\begin{equation}\label{eq:continuity:lagrangean}
  \frac{D\rho}{D t} = -\rho(\bigtriangledown \cdot v).
\end{equation}
%
In the free homologous expansion stage (${v \propto r}$),
  equation~(\ref{eq:continuity:lagrangean}) is simplified:
\begin{equation}\label{eq:continuity:lagrangean:homogeneity}
  \frac{D\rho}{D t} + \frac{3\rho}{t} = 0.
\end{equation}
In this case, the rate of transitions to any discrete bound or free level $i$
  of neutral or ionized hydrogen can be written as
\begin{equation}\label{eq:kinetics}
  \frac{Dn_{i}}{D t} + \frac{3n_{i}}{t} = K_{i}(t),
\end{equation}
  where
  $n_{i}$ is the population of level $i$ for an atom or ion.
Neglecting the processes of induced emission
  for the ground level of hydrogen,
  we define the function $K_1(t)$ as
\begin{equation}\label{eq:kinetics:K:H0}
  K_{1}(t) = (N(t) - n_1 - \WI{n}{e})\;(A_{2q} + A_{21}) + n_1B_{12}J_{12}(t),
\end{equation}
Here
\begin{equation}
  N(t) = N_0\left(\frac{t_0}{t}\right)^3
\end{equation}
  is the hydrogen number density,
  and $A_{2q}$ is the two-photon decay probability ${2\mathrm{s} \to 1\mathrm{s}}$.
The reverse ${1\mathrm{s} \to 2\mathrm{s}}$
  transition (two-photon absorption) rate
  is much lower than the ${2\mathrm{s} \to 1\mathrm{s}}$ rate,
  and we neglect this process
  \citep{PotashovBlinnikovUtrobin2017a};
  we also neglect stimulated two-photon decays.
$A_{21}$ and $B_{12}$ are the Einstein coefficients for spontaneous and induced
  transition ${1 \leftrightarrow 2}$,
  and $J_{12}(t)$ is the mean intensity of radiation for the ${2 \to 1}$ transition
  averaged over the line profile.

Equation (\ref{eq:kinetics}) with the index
  $\mathrm{e}$ instead of $i$ is valid
  for the number density of free electrons,
  with
\begin{equation}\label{eq:kinetics:K:H1}
  \WI{K}{e}(t) = (N(t) - n_1 - \WI{n}{e})\;P_{2c}(t) - \WI{n}{e}^2\;R_{c2}(t)\;.
\end{equation}
Here $P_{2c}(t)$ is the photoionization coefficient for the second level,
  and $R_{c2}(t)$ is the radiative recombination coefficient for the second level.

In our model, we will use the fact that the main contribution to the opacity
  in the frequency band of the Lyman continuum
  ${\nu \geqslant \nu_{LyC}}$
  is provided by free-bound processes
  \citep{PotashovBlinnikovUtrobin2017a}.
We neglect the relatively small contribution
  from bound-bound processes (expansion opacity)
  and free-free processes
  in the emission and absorption coefficients.
The absorption in this band is mainly due to neutral hydrogen, and the optical depth is very large.
Therefore, there is virtually no photospheric radiation,
  and the radiation field for the regions above the photosphere
  is determined by the diffusive radiation.
In this case, it can be shown (see Appendix~\ref{ap:balance})
  that the rates of photoionization transitions from the ground level of hydrogen
  and recombination to the ground level completely coincide
  (even if there is no pure hydrogen in the SN envelope).
Thus, the ground level of hydrogen is in detailed balance with the continuum,
  and the related processes are not included in the system of equations
  (\ref{eq:kinetics:K:H0} and \ref{eq:kinetics:K:H1}).

Let us assume that the population of the second level is relatively small,
\begin{equation}\label{eq:n2_small}
  n_2 = N(t) - n_1 - \WI{n}{e} \ll n_1,
\end{equation}
  and write down the standard formulas for the Sobolev approximation
  \citep{Sobolev1960book, Castor1970}.

The Sobolev optical depth in L$\alpha$ is
\begin{equation}
  \tau_S(t) \simeq \frac{c^3}{8\pi}\frac{A_{21}}{\nu_{L\alpha}^3}\frac{g_2}{g_1}n_1t.
\end{equation}
The frequency-averaged mean intensity of the transition ${1 \leftrightarrow 2}$ is
\begin{equation}\label{eq:jlu}
  J_{12}(t) = (1 - \beta(t)) \; S(t) + \beta(t) \, J_c(\nu_{L\alpha}, t),
\end{equation}
  where $J_c(\nu_{L\alpha},t)$ is the average intensity of the continuum
  on the L$\alpha$ frequency.

Because the medium is optically thick for L$\alpha$ photons, ${\tau_S(t) \gg 1}$.
Then the escape probability of L$\alpha$ photons,
  summed over all directions and line frequencies, is
\begin{equation}\label{eq:beta}
  \beta(t) = \frac{1-e^{-\tau_S(t)}}{\tau_S(t)} \simeq \frac{1}{\tau_S(t)}.
\end{equation}
The source function is
\begin{equation}\label{eq:slu}
  S(t) \simeq \frac{2 h \nu_{L\alpha}^3}{c^2}
    \left( \frac{g_1 n_2}{g_2 n_1} \right).
\end{equation}
All other notations are standard.

It is known that the optical depth of the SN~II envelope
  in the Lorentz wings is very large for L$\alpha$
  (${a \tau_S \gg 1}$, where $a$ is the Voigt damping parameter),
  and the profile cannot be considered as the Doppler one.
In this case, under the hypothesis of complete frequency redistribution,
  the formal criterion of the applicability of Sobolev theory is violated
  \citep{Chugai1980a}.
However, \citet{Chugai1980a} and \citet{Grachev1989}
  have shown that the estimation (\ref{eq:beta}) remains true
  for the conservative scattering of L$\alpha$ photons
  if one assumes partial frequency redistribution
  and neglects the effects of recoil.
In these works, the Fokker-Planck approximation was used
  for the redistribution function.
\citet{HummerRybicki1992} showed
  that the frequency-weighted mean intensity $J_{12}(t)$ defined in (\ref{eq:jlu}) is
  weakly dependent on the mechanism of frequencies redistribution.
Expressions (\ref{eq:jlu}) and (\ref{eq:beta}) remain correct
  in the case of non-conservative scattering with partial frequencies redistribution
  and partial non-coherence owing to the Stark effect
  \citep{Chugai1988a}.

When absorption in the continuum in the region of a line is significant
  (for example, L$\alpha$ photons can ionize Ca~II from the second level),
  one must take into account corrections to the Sobolev approximation
  \citep{HummerRybicki1985, Chugai1987, Grachev1988}.
The additional selective absorption in lines
  of metal admixtures may also play a role,
  because a large number of lines of Fe~II and Cr~II are
  in the vicinity of L$\alpha$
  \citep{Chugai1988b, Chugai1998}.
These processes increase the value of the effective escape probability
  (\ref{eq:beta}) (the so-called loss probability of a photon in flight).
The influence of absorption in the continuum
  and in the lines of admixtures
  on the time-dependence effect was investigated by
  \citet{PotashovBlinnikovUtrobin2017a}.
In the current analytical approach, we use a simple model and
  do not take into account these processes,
  using hereafter equations (\ref{eq:jlu}) and (\ref{eq:beta}).
However, in Section~\ref{sec:full_system},
  these modifications of the Sobolev approximation
  will be taken into account numerically.

Combining
(\ref{eq:kinetics}--\ref{eq:kinetics:K:H1},
  \ref{eq:jlu}--\ref{eq:slu}),
  we obtain the system
\begin{equation*}
  \dot{n}_{1} =
    (N(t) - n_1 - \WI{n}{e})\;(A_{2q} + A_{21}\beta(t))
    -
    n_1B_{12}\beta(t) J_c(\nu_{L\alpha}, t) - \frac{3n_1}{t},
\end{equation*}
\begin{equation*}
  \WI{\dot{n}}{e} = (N(t) - n_1 - \WI{n}{e})\;P_{2c}(t) - \WI{n}{e}^2\;R_{c2}(t) - \frac{3\WI{n}{e}}{t}.
\end{equation*}
In accordance with
  \citet{Mihalas1978book}
  and
  \citet[p.~273]{HubenyMihalas2014book},
  the total photoionization coefficient for the second level is the integral
\begin{equation}\label{eq:p2c}
  P_{2c}(t) = 4\pi\int\limits_{\nu_2}^{\infty} J_c(\nu, t)\frac{\alpha_{2c}(\nu)}{h\nu}d\nu,
\end{equation}
  where $\nu_2$ is a frequency of the ${2 \leftrightarrow c}$ transition,
  $c$ means the continuum,
  and $\alpha_{2c}$ is a photoionization cross-section of the second level
  at the frequency $\nu$.
The total radiative recombination coefficient
  of the second level for a purely hydrogen plasma
  looks like
\begin{equation}\label{eq:rc2}
  R_{c2}
    = 4\pi\,\Phi_{\mathrm{Saha}}(\WI{T}{e})\;\int\limits_{\nu_2}^{\infty} \frac{\alpha_{2c}(\nu)}{h\nu}\frac{2h\nu^3}{c^2}e^{-\frac{h\nu}{k\WI{T}{e}}}d\nu
    \approx
    \frac{64 \pi^5 m e^{10}}{3 \sqrt{3}\,c^3\,h^6}\,g_{\,\mathrm{II}}(2,\nu_{2})\,\Phi_{\mathrm{Saha}}(\WI{T}{e})\,E_1\left(\frac{h\nu_2}{k\WI{T}{e}}\right)
\end{equation}
  if the processes of induced emission are neglected.
%
Here
  $\Phi_{\mathrm{Saha}}(\WI{T}{e})$ is the Saha-Boltzmann factor;
  $g_{\,\mathrm{II}}(2,\nu_{2})$ is a bound-free Gaunt factor for the ${2 \leftrightarrow c}$ transition;
  and $E_1$ is the exponential integral.
Note that $R_{c2}$ is constant over time if $\WI{T}{e}$ is constant.
It would be more realistic
  to use the effective coefficient of recombination to the second level
  calculated for a multilevel atom taking into account cascade transitions
  instead of the direct  coefficient of recombination to the same level (\ref{eq:rc2}).
Usually, Lyman lines
  are supposed to be optically thick,
  and the escape probabilities for
  the Lyman line photons are zero,
  while all other lines are optically thin.
This is the so-called Case~B
  \citep{BakerMenzel1938, OsterbrockFerland2006book}
  for the recombination coefficient calculation.
In the opposite Case~A
  \citep{BakerMenzel1938},
  full transparency is assumed
  and the escape probabilities for the Lyman line
  photons are equal to unity.
In reality, the escape probability of a photon in Lyman line
  differs from both zero and unity.
Therefore, neither Case A nor Case B is fully implemented,
  and the real situation is intermediate
  \citep{HummerStorey1987, HummerStorey1992}.
We choose the average value of $R_{c2}$ from \citet{Hummer1994}
  for our simple system.

Let us now introduce the dimensionless variables
\begin{equation*}
  u_1(t) = \frac{n_1}{N(t)} = \frac{n_1}{N_0}\frac{t^3}{t_0^3},
  \qquad
  \WI{u}{e}(t) = \frac{\WI{n}{e}}{N(t)} = \frac{\WI{n}{e}}{N_0}\frac{t^3}{t_0^3}
\end{equation*}
  that are normalized to the full current number density.
By rewriting the system we obtain
\begin{equation}\label{eq:kinetics:td:u1}
  \dot{u}_{1} = (1-u_1-\WI{u}{e})\;
    \left[A_{2q} + \frac{\tilde{A}}{u_1}
    \left(\frac{t}{t_0}\right)^2
    \right]
    -
    \tilde{B} J_c(\nu_{L\alpha}, t) \left(\frac{t}{t_0}\right)^2,
\end{equation}
\begin{equation}\label{eq:kinetics:td:ue}
  \WI{\dot{u}}{e} = (1-u_1-\WI{u}{e})\;P_{2c}(t)-\WI{u}{e}^2\tilde{R}\left(\frac{t_0}{t}\right)^3.
\end{equation}
Here we also introduce a new notation for constants:
\begin{equation}\label{eq:abr}
  \tilde{A} {=} \frac{8 \pi \nu_{L\alpha}^3}{c^3} \frac{g_1}{g_2} \frac{1}{N_0 t_0},\quad
  \tilde{B} {=} \frac{4\pi}{h\,c} \frac{1}{N_0 t_0},\quad
  \tilde{R} {=} N_0 R_{c2}.
\end{equation}
Especially important for further simplification of the system
  (\ref{eq:kinetics:td:u1}, \ref{eq:kinetics:td:ue})
  is an investigation of the behaviour of $J_c(\nu_{L\alpha}, t)$ and $P_{2c}(t)$
  over time.
The continuum radiation between the Lyman and Balmer edges
  is assumed to be \emph{given},
  and not determined from the self-consistent calculation.

In the optically thin case we can write,
  introducing a dilution factor $W(t)$,
\begin{equation}\label{eq:jc:thin}
  J_c(t) = W(t) B(T_c),
\end{equation}
If we additionally assume that the region under consideration is sufficiently
  far from the photosphere, the dilution factor is
\begin{equation}\label{eq:dilution_factor}
  W(t) \simeq \frac{1}{4}\left(\frac{\WI{R}{ph}}{Vt}\right)^2.
\end{equation}
Then the continuum intensity $J_c(\nu_{L\alpha}, t)$
  and the photoionization coefficient $P_{2c}(t)$
  decrease as ${\sim 1/t^2}$.
This case is called the free-streaming approximation.

In a real SN, the medium at the considered frequencies
  in the continuum is optically thick.
Numerous metal lines between
  the Lyman and Balmer ionization thresholds contribute to
  the large expansion opacity
  for SN envelopes with non-zero metallicities.
The averaged intensity of such a quasi-continuum
  is lower than in the optically thin limit.
Even in this case, numerical simulation
  (for example using the \code{STELLA} code)
  shows the power dependence of the intensity
  and the photoionization coefficient on time.
Namely,
\begin{equation}\label{eq:jc}
  J_c(\nu_{L\alpha}, t) \simeq J_c(\nu_{L\alpha}, t_0) \left(\frac{t_0}{t}\right)^{s_1},
\end{equation}
\begin{equation}\label{eq:p2c2}
  P_{2c}(t) \simeq P_{2c}(t_0) \left(\frac{t_0}{t}\right)^{s_2} = \tilde{P} \left(\frac{t_0}{t}\right)^{s_2}.
\end{equation}
The values of $s_1$ and $s_2$ depend on the distance from the photosphere,
  but they are always greater than 2.
In the general case, we limit the domain of these exponents to
  ${s_1 \geqslant 2}$ and ${s_2 \geqslant 2}$.
In the optically thick case, the
  typical values of these exponents can be significant
  (see Table~\ref{tab:values}, Appendix~\ref{ap:values}).

Now let us introduce the normalized time
\begin{equation}\label{eq:time}
  \tau = \frac{t}{t_0}.
\end{equation}
Taking into account
   (\ref{eq:jc}), (\ref{eq:p2c2}) and (\ref{eq:time}),
  we can rewrite the system (\ref{eq:kinetics:td:u1}, \ref{eq:kinetics:td:ue}) as
\begin{equation}\label{eq:kinetics:td2:u1}
  \dot{u}_{1} = (1-u_1-\WI{u}{e})\Big(Q + \frac{A}{u_1}\tau^2\Big)
  -\frac{B}{\tau^{s_1-2}},
\end{equation}
\begin{equation}\label{eq:kinetics:td2:ue}
  \WI{\dot{u}}{e} = (1-u_1-\WI{u}{e})\frac{P}{\tau^{s_2}}
    -\WI{u}{e}^2\frac{R}{\tau^3}.
\end{equation}
  Now $u_1$ and $\WI{u}{e}$ are functions of $\tau$, and
\begin{equation}\label{eq:abr2}
  Q = A_{2q}\,t_0,\quad
  A = \tilde{A}\,t_0,\qquad
  B = \tilde{B}J_c(\nu_{L\alpha}, t_0)\,t_0,\qquad
  P = \tilde{P}\,t_0,\qquad
  R = \tilde{R}\,t_0.
\end{equation}
Table~\ref{tab:values} shows typical values
  of the constants
  $Q$, $A$, $B$, $P$, $R$, $s_1$, $s_2$
  that correspond to the physical conditions for the outer, high-velocity layers
  located far from the photosphere (column 1),
  and the near-photospheric layers (column 2) of SNe~IIP,
  (see Appendix~\ref{ap:values} for details).

The initial conditions for the problem are:
\begin{equation}\label{eq:kinetics:td2:init}
  \begin{split}
    &0 < u_1(1) \leqslant 1,
    \quad
    0 \leqslant \WI{u}{e}(1) \leqslant 1,\\
    &
    u_2(1) = 1 - u_1(1) - \WI{u}{e}(1) \ll u_1(1)
    \ \ (\mbox{see eq.~}\ref{eq:n2_small}).
  \end{split}
\end{equation}
\begin{table}
  \centering
  \caption{
  Typical values of dimensionless constants for the system
    (\ref{eq:kinetics:td2:u1}, \ref{eq:kinetics:td2:ue})
    obtained on the basis of \code{STELLA} calculations for SN~1999em
    \citep{BaklanovBlinnikovPavlyuk2005, PotashovBlinnikovUtrobin2017a},
    which is a standard SN~IIP.
  The column labelled ``Out'' contains values
    that correspond to the outer, high-velocity layers located far from the photosphere.
  The column labelled ``Ph'' contains values that correspond to the near-photospheric layers.
  See Appendix~\ref{ap:values} for details.}
  \label{tab:values}
  \renewcommand{\arraystretch}{1.5}
  \begin{tabular}{ccc}
    \toprule
    &
    out &
    ph \\
    \midrule
    $Q$         & $3.5\cdot10^6$ & $3.5\cdot10^6$ \\
    $A$         & $3\cdot10^7$ & $3\cdot10^4$ \\
    $B$         & $6\cdot10^{-4}$ & $6\cdot10^{-3}$ \\
    $P$         & $7\cdot10^{9}$ & $2\cdot10^{11}$ \\
    $R$         & $50$ & $5\cdot10^{4}$ \\
    $s_1$       & $21$ & $24$ \\
    $s_2$       & $4$ & $6$ \\
    \bottomrule
  \end{tabular}
\end{table}

Equilibrium populations in the same approximation can be found by solving a system
  of algebraic equations:
\begin{equation}\label{eq:kinetics:ss:u1}
  (1-u_1^{ss}-\WI{u}{e}^{ss})\Big(Q + \frac{A}{u_1^{ss}}\tau^2\Big)
  -\frac{B}{\tau^{s_1-2}} = 0,
\end{equation}
\begin{equation}\label{eq:kinetics:ss:ue}
  (1-u_1^{ss}-\WI{u}{e}^{ss})\frac{P}{\tau^{s_2}} - (\WI{u}{e}^{ss})^2\frac{R}{\tau^3} = 0,
\end{equation}
  where the superscript $ss$ denotes steady state.
Thus, the answer to the question regarding the importance of the time-dependence effect
  in kinetics can be found by comparing the solutions of the systems
  (\ref{eq:kinetics:td2:u1}, \ref{eq:kinetics:td2:ue})
  and (\ref{eq:kinetics:ss:u1}, \ref{eq:kinetics:ss:ue}).
In the next section, it will be shown that at large time
  the deviation of $u_1$, $\WI{u}{e}$ from $u_1^{ss}$, $\WI{u}{e}^{ss}$
  persists regardless of the initial conditions.

\section{The system at large time}
  \label{sec:large_time}

We will investigate the solutions of systems
  (\ref{eq:kinetics:td2:u1}, \ref{eq:kinetics:td2:ue}) and
  (\ref{eq:kinetics:ss:u1}, \ref{eq:kinetics:ss:ue})
  at large time ${\tau \to \infty}$.
In this case, we will assume that the photospheric phase lasts for an indefinitely
  long time.

Let us transform the original system
  (\ref{eq:kinetics:td2:u1}, \ref{eq:kinetics:td2:ue})
  by introducing the function $u_2$
  according to the relationship ${u_1{+}u_2{+}\WI{u}{e}{=}1}$.
We should keep only terms linear in $u_2$
  because the smallness of this function was already assumed in the derivation
  (see~\ref{eq:n2_small} and \ref{eq:kinetics:td2:init}).
Thus, this is a reduction of the source system to a ``normal'' form:
\begin{align}
  \dot{u}_1&=u_2 g_1(\tau,u_1)-g_2(\tau),\label{eq:kinetics:td3:u1}
  \\
  \dot{u}_2&=-u_2 g_3(\tau,u_1)+g_4(\tau,u_1),\label{eq:kinetics:td3:u2}
\end{align}
  where we have introduced the following functions:
\begin{align*}
  &g_1(\tau,u_1) = Q + \frac{A}{u_1} \tau^2,\\
  &g_2(\tau) = \frac{B}{\tau^{s_1-2}},\\
  &g_3(\tau,u_1) = Q + \frac{A}{u_1} \tau^2+\frac{P}{\tau^{s_2}}
    +2\,(1{-}u_1)\frac{R}{\tau^3},\\
  &g_4(\tau,u_1) = \frac{B}{\tau^{s_1-2}}
    +(1{-}u_1)^2\frac{R}{\tau^3}.
\end{align*}
It is important that, because of its linearity, equation (\ref{eq:kinetics:td3:u2})
  can be integrated explicitly and its solution can be written as
\begin{equation}\label{eq:kinetics:td3:u2:sol}
  u_2(\tau) = 
    \!e^{-G_3(\tau)}\Big[u_2(1)+\int\limits_{1}^{\tau}g_4(\tau'\!\!,u_1(\tau'))e^{G_3(\tau')}d\tau'\Big],
\end{equation}
  where
\begin{equation}\label{eq:G3}
  G_3(\tau)\equiv\int\limits_{1}^{\tau}g_3(\tau',u_1(\tau'))d\tau'.
\end{equation}
Equation (\ref{eq:kinetics:td3:u2:sol}) together with (\ref{eq:G3})
  represents a formal solution for $u_2$, because $u_1$ is still unknown.

Analysis of the system (\ref{eq:kinetics:td3:u1}, \ref{eq:kinetics:td3:u2})
  shows that the solutions that satisfy the initial conditions
  $0 {<} u_1(1) {\leqslant} 1$,
  $0 {\leqslant} u_2(1) {\leqslant} 1$
  remain bounded
  and stable in the sense of Lyapunov
  \multicitep{Demidovich1967book, p.~66; Khalil2002book, p.~111},
  see also Appendix~\ref{ap:tube}.
The functions $g_1(t)$ and $g_3(t)$
  increase rapidly,
  ${g_{1,3} \sim \mathcal{O}(\tau^2)}$,
  if $u_1$ is bounded and finite.
In contrast, the functions $g_{2,4}$ are decreasing
  as some power of $\tau$.
We can use the fact that
  for any bounded behaviour of $u_1(\tau)$,
  $G_3$ is a rapidly increasing function of time,
  at least as ${\sim \mathcal{O}(\tau^3)}$.
Therefore, terms with $\exp(G_3)$ are most important at large time.
The term with $u_2(1)$ in (\ref{eq:kinetics:td3:u2:sol})
  is exponentially small at large time;
  that is, $u_2$ ``forgets'' the initial conditions.
After multiple integration of the second term by parts
  and neglecting the exponentially small terms
  at ${\tau\to \infty}$, we obtain
\begin{equation}\label{eq:kinetics:td3:u2:sol_approx}
  u_2\simeq\frac{g_4(\tau,u_1)}{g_3(\tau,u_1)}-\frac{1}{g_3(\tau,u_1)}\frac{d}{d\tau}\!\!\left(\frac{g_4(\tau,u_1)}{g_3(\tau,u_1)}\right)+\dots
\end{equation}
The first term of this decomposition is actually a steady-state approximation
  of equation (\ref{eq:kinetics:td3:u2}).
Substituting it into (\ref{eq:kinetics:td3:u1}), we obtain the equation for $u_1$:
\begin{equation}\label{eq:kinetics:td4:u1}
  \dot{u}_1 = f(\tau,u_1) = \frac{g_1(\tau,u_1)g_4(\tau,u_1)}{g_3(\tau,u_1)}
    -g_2(\tau).
\end{equation}
Equation (\ref{eq:kinetics:td4:u1})
  is a special case of the Appell equation \citep{Appell1889}.
That is, it is a generalized Abel equation of the second kind
  \citep{PolyaninZaitsev2002book, Semenov2014}
  and unfortunately it is generally not integrable by quadratures.

\begin{figure}
  \centering
  \begin{minipage}{0.5\textwidth}
    \includegraphics[width=\textwidth]{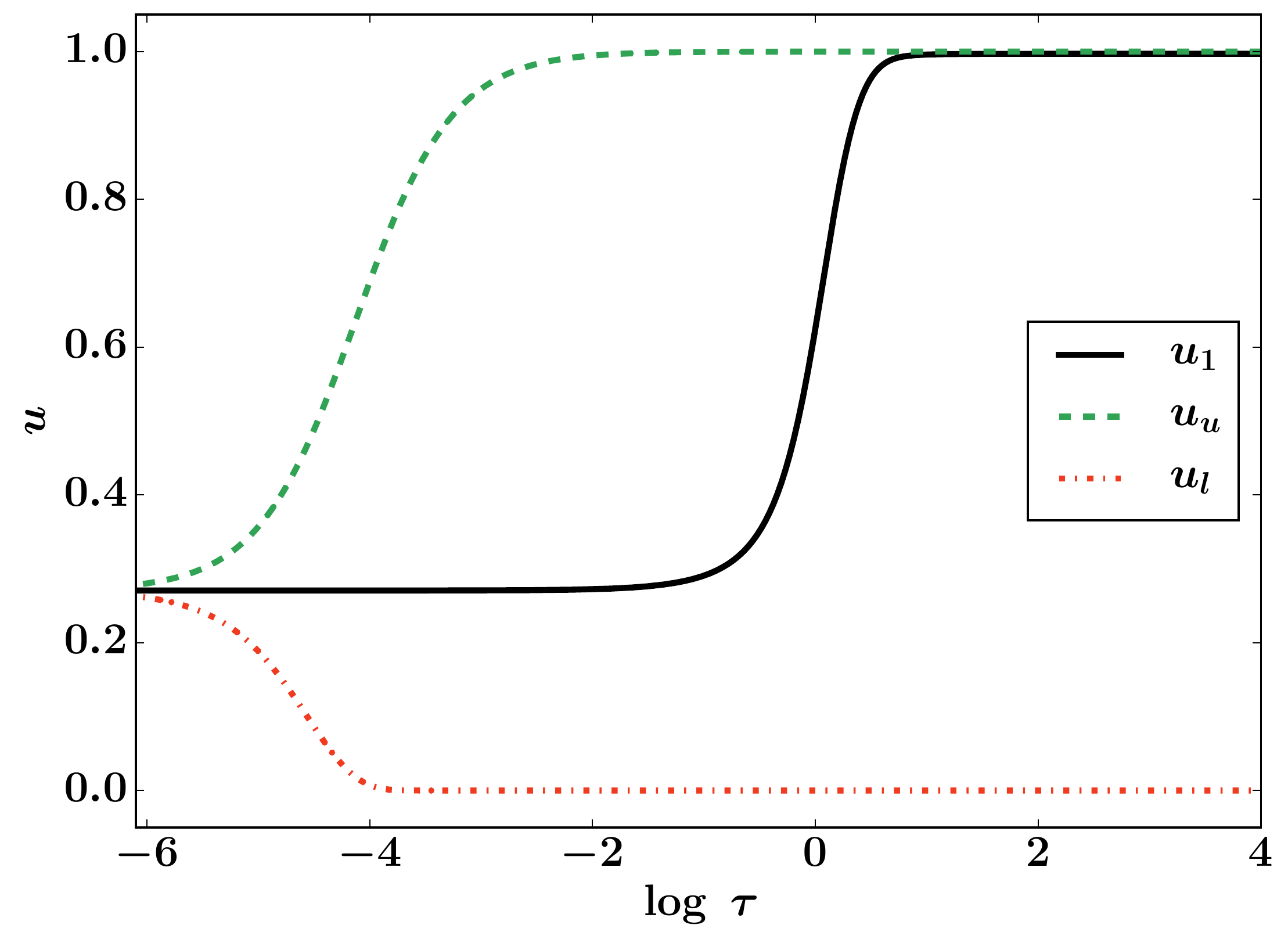}
  \end{minipage}
  \caption {
  The population of the first level $u_1(\tau)$
    and the bounding solutions $u_u(\tau)$, $u_l(\tau)$
    calculated with physical parameters typical
    for the layers located near the photosphere
    of SN~1999em
    (see Table~\ref{tab:values} column~2 and Appendix~\ref{ap:values}).
  }
  \label{fig:u1_uu_ul}
\end{figure}

Usually, approximate analytical methods, such as
  the small parameter method \citep[p.~405]{Fedoruk1985book},
  Chaplygin method \citep[p.~260]{BerezinZhidkov1959book}
  or power series method,
  are used to solve nonlinear differential equations.
Case (\ref{eq:kinetics:td4:u1}) requires
  \emph{a large} number
  of steps in each particular method
  or many summands in the decomposition.
The analytical approach becomes cumbersome and useless.
But we do not need to solve (\ref{eq:kinetics:td4:u1}).
It is sufficient to prove that any solution of this equation
  that satisfies the initial conditions ${0 {<} u_1(1) {\leqslant} 1}$
  is bounded in the interval ${(0, 1)}$.
To do this, consider two functions
\begin{equation}\label{eq:f_l}
  f_l(\tau,u) = - |u|\,\frac{P B}{A\,\tau^{s_1+s_2}},
\end{equation}
\begin{equation}\label{eq:f_u}
  f_u(\tau,u) = (1{-}u)^2\frac{R}{\tau^3}.
\end{equation}
Let us take an arbitrary moment of time ${\tau_1 \gg 1}$.
In the domain (${\tau \in [1, \tau_1];\;u \in \mathbb{R}}$),
  the inequalities $f_l(\tau,u) \leqslant f(\tau,u) \leqslant f_u(\tau,u)$ are obviously satisfied.
Because the function $f(\tau,u)$ is continuously differentiable
  in this domain and satisfies the Lipschitz condition,
  according to the Chaplygin theorem on differential inequalities
  \multicitep{BerezinZhidkov1959book, p.~260; Khalil2002book, p.~102},
  one can write ${u_l(\tau) \leqslant u_1(\tau) \leqslant u_u(\tau)}$
  for any time in ${[1, \tau_1]}$,
  where $u_l$, $u_u$ are solutions of the equations
  ${\dot{u}_l = f_l(\tau,u_l)}$,
  ${\dot{u}_u = f_l(\tau,u_u)}$, respectively.
Their initial conditions must match the initial conditions of
  (\ref{eq:kinetics:td4:u1})
  ${u_1(1) = u_l(1) = u_u(1) = u_0}$.
These solutions provide bounds for the function $u_1(\tau)$
  (Fig.~\ref{fig:u1_uu_ul}),
  according to Chaplygin they are called ``barrier'' solutions.
The following expressions complete the proof:
\begin{equation}\label{eq:ul_solution}
  u_l(\tau) = u_0\,\exp\left[{-}\frac{B P\,(1{-}\tau^{-(s_1{+}s_2{-}1)})}{A (s_1+s_2-1)}\right] > 0,
\end{equation}
\begin{equation}\label{eq:uu_solution}
  u_u(\tau) = 1-
    \left[
    \frac{1}{1-u_0}-\frac{R}{2}\left(1{-}\tau^{-2}\right)
    \right]^{-1} < 1.
\end{equation}

\begin{figure}
  \centering
  \begin{minipage}{0.5\textwidth}
    \includegraphics[width=\textwidth]{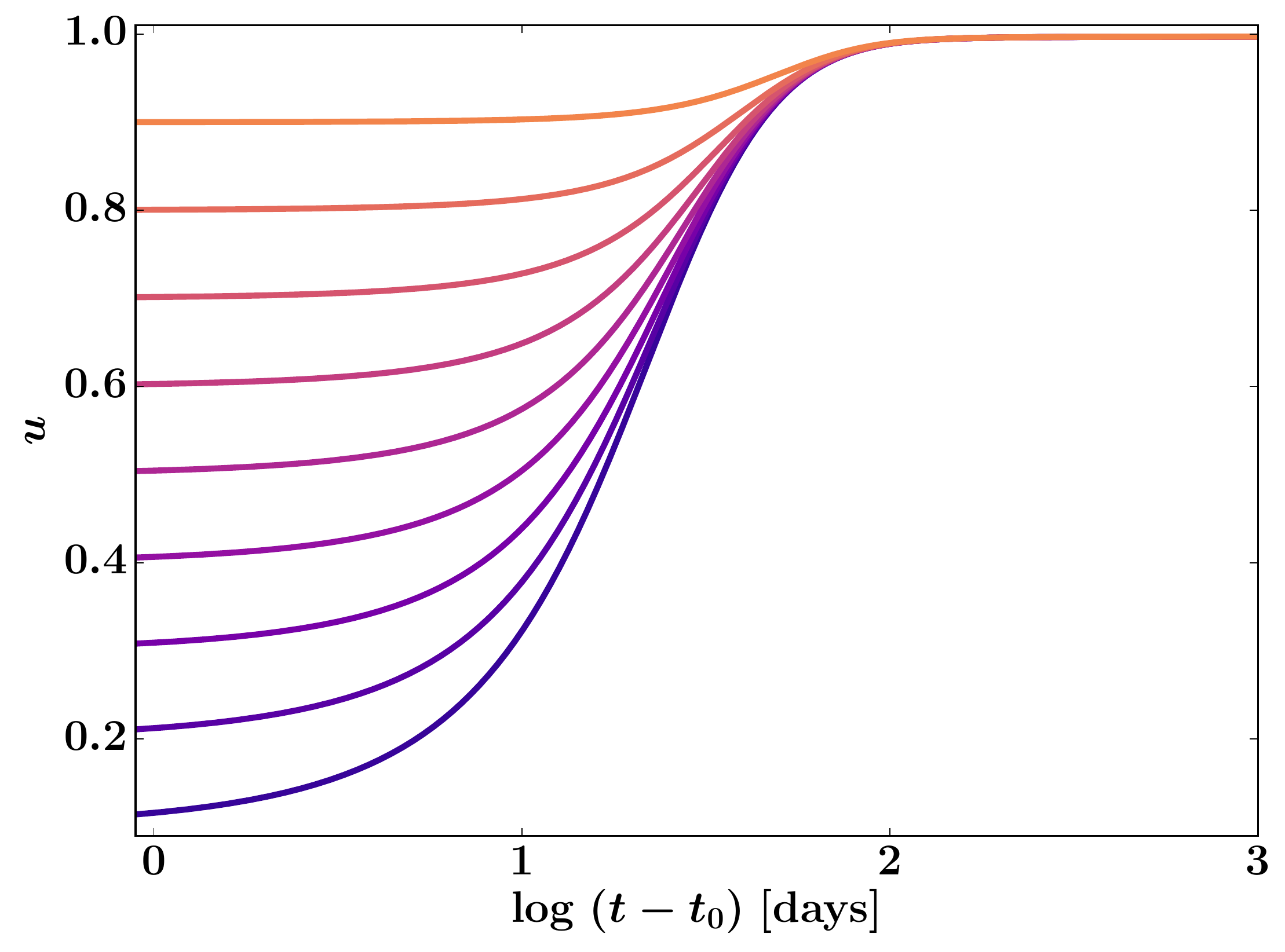}
  \end{minipage}
  \caption {
  The dissipative behaviour of $u_1(t)$ functions with different initial conditions
    for physical parameters typical of
    the near-photospheric optically thick layers
    of SN~1999em
    (see Table~\ref{tab:values} column~2, Appendix~\ref{ap:values}).
  Here $t$ is a physical time and ${t_0 {=} 20}$~d.
  }
  \label{fig:tube}
\end{figure}

In addition, we give another proof that points
  to an interesting property of equation (\ref{eq:kinetics:td4:u1}).
The function $f(u_1,\tau)$ in the considered range of $u_1$
  decreases strictly monotonically, and
\begin{align}
  f(\tau,0)&=\frac{R}{\tau^3} > 0, \\
  f(\tau,1)&=-\frac{B P\, \tau^{-(s_1+s_2-2)}}
    {Q + A \tau^2+P/\tau^{s_2}} < 0.
\end{align}
These boundary properties do not allow the function $u_1(\tau)$
  to go beyond the interval ${(0, 1)}$.
In addition, ${f(\tau,u_m)<f(\tau,u_n)}$
  holds for any ${u_m>u_n}$
  owing to the monotonicity of $f(\tau,u)$.
Therefore,
\begin{equation}\label{eq:dumn}
  \frac{d(u_m{-}u_n)}{d \tau} = f(\tau,u_m) - f(\tau,u_n) < 0.
\end{equation}
Equation (\ref{eq:dumn}) guarantees the convergence of any two solutions $u_m$ and $u_n$.
Their difference is continuously decreasing and never changes sign
  because two different solutions of equation (\ref{eq:kinetics:td4:u1})
  cannot intersect according to Cauchy's theorem on existence and uniqueness.
  \multicitep{Fedoruk1985book, p.~10; Khalil2002book, p.~88}.
Thus, any solution of equation (\ref{eq:kinetics:td4:u1})
  starting from arbitrary time
  eventually lies inside a cylinder of non-zero radius
  and never goes out of it (Fig.~\ref{fig:tube}).
This property is called dissipativity, and systems that
  demonstrate it, for example the original system
  (\ref{eq:kinetics:td2:u1}, \ref{eq:kinetics:td2:ue}),
  are called a dissipative systems
  \multicitep{Demidovich1967book, p.~287; Khalil2002book, p.~168}.
Fig.~\ref{fig:tube} shows that the convergence time
  of the solutions of the dissipative system to the cylinder-tube is large
  (of the order of 10~d for the typical SN~IIP conditions).
It turns out that if one takes collisional processes into account,
  the convergence time decreases to only thousands of seconds.
Moreover, if one neglects the width of the cylinder-tube, it is possible to say that the system ``forgets'' the initial conditions!
See Appendix~\ref{ap:tube} for details.

From (\ref{eq:kinetics:td4:u1}) it follows that
  ${\lim_{\tau\to\infty} \dot{u}_1 = 0}$.
The function $u_1(\tau)$ becomes constant
  ${u_1(\tau{=}\infty)}$ at large time $\tau{=}\infty$.
Because $u_l(\tau)$ and $u_u(\tau)$ bound the function $u_1(\tau)$,
  it follows from (\ref{eq:ul_solution}, \ref{eq:uu_solution})
  that ${0 < u_1(\tau{=}\infty) < 1}$.
From (\ref{eq:kinetics:td3:u2:sol_approx}), it also follows
  that ${u_2 \to 0}$ at ${\tau \to \infty}$.
This means that the real normalized electron number density
  becomes constant ${0 <\WI{u}{e}(\tau {=} \infty) < 1}$
  at large time, regardless of the initial conditions.
But by solving (\ref{eq:kinetics:ss:u1}, \ref{eq:kinetics:ss:ue})
  it can be shown that the equilibrium normalized electron number density
  approaches zero as
\begin{equation}\label{eq:kinetics:ss:ue:power}
  \WI{u}{e}^{ss} \sim \tau^{-(s_1 + s_2 - 3)/2}.
\end{equation}
Its clear that in the time-dependent case
  the envelope expands with a greater degree of ionization
  compared with the steady-state solution.
A similar phenomenon is observed
  in atmospheric explosions
  \multicitep{Raizer1959; ZeldovichRajzer2008book} and
  in the early Universe under cosmological conditions
  with a slowing down of the recombination of the primordial plasma
  \citep{ZeldovichKurtSunyaev1969, Peebles1968, KurtShakhvorostova2014a}.
The number density of free electrons experiences a ``freeze-out'' in this case.
Unlike the ``freeze-out'' effect in terrestrial atmospheric explosions,
  the effect of time-dependence in SNe
  remains important even when the temperatures of material and radiation are \emph{constant}.
This is true, for example, for the optically thin case
  ${s_1=2}$, ${s_2=2}$ under the free-streaming approximation.

\begin{figure*}
  \begin{minipage}{0.47\textwidth}\centering
    \includegraphics[width=\textwidth]{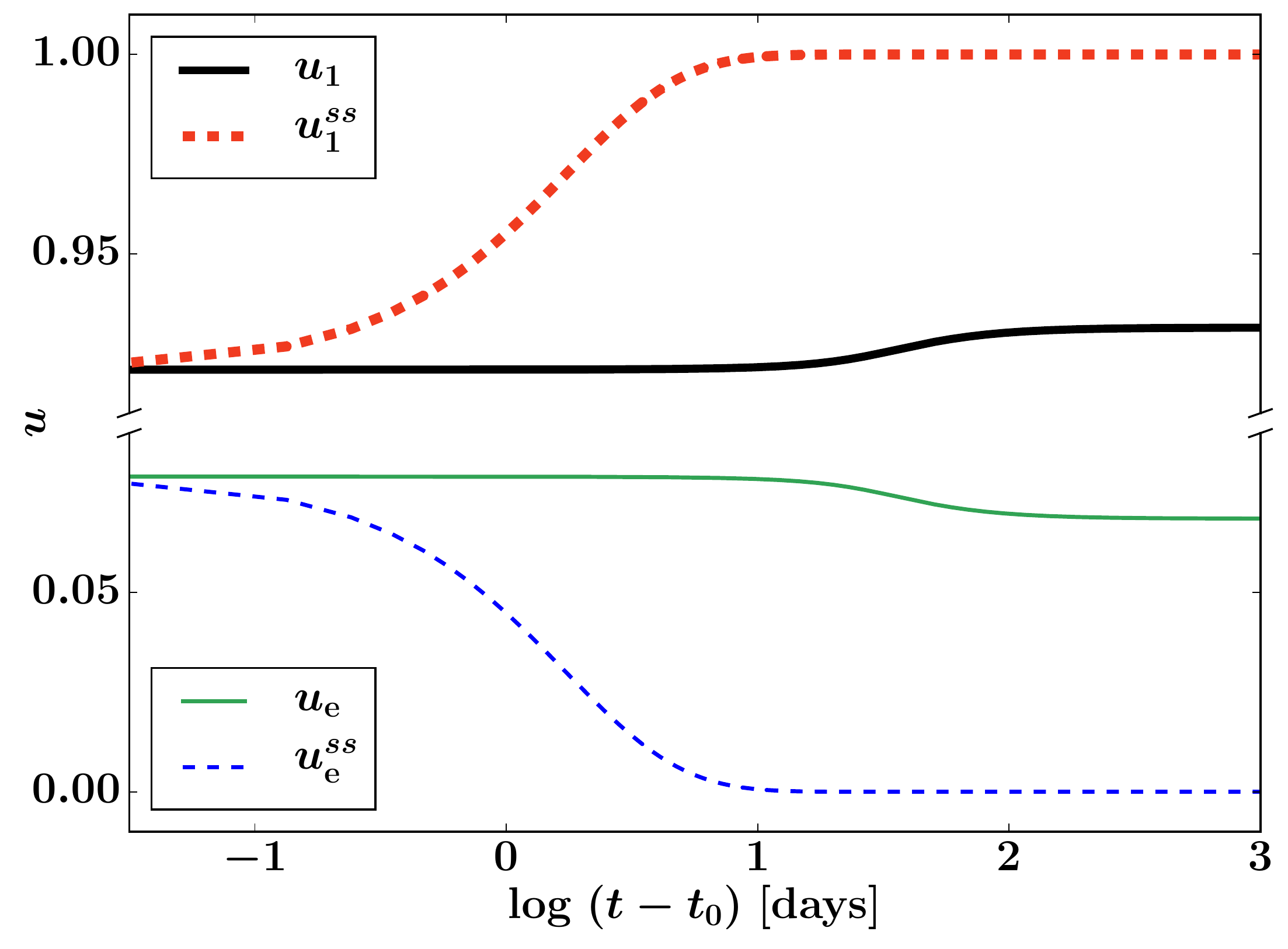}
    \caption {
    Number density functions
      $u_1(t)$, $u_1^{ss}(t)$, $\WI{u}{e}(t)$, and $\WI{u}{e}^{ss}(t)$
      depending on physical time ($t_0 = 20$~d),
      calculated for the physical parameters typical for the layers
      located far beyond the photosphere
      of SN~1999em
      (see Table~\ref{tab:values} column~1, Appendix~\ref{ap:values}).
    Initial conditions are ${u_1(t_0) = u_1^{ss}(t_0)}$ and
      $\WI{u}{e}(t_0) = \WI{u}{e}^{ss}(t_0)$.
    The normalized electron number density $\WI{u}{e}(t)$
      becomes constant $\WI{u}{e}(t{=}\infty) \approx 0.05$ at large time.
    Note the break in the vertical axes.
    }
    \label{fig:u_ss_out}
  \end{minipage}
  \hfill
  \begin{minipage}{0.47\textwidth}\centering
    \includegraphics[width=\textwidth]{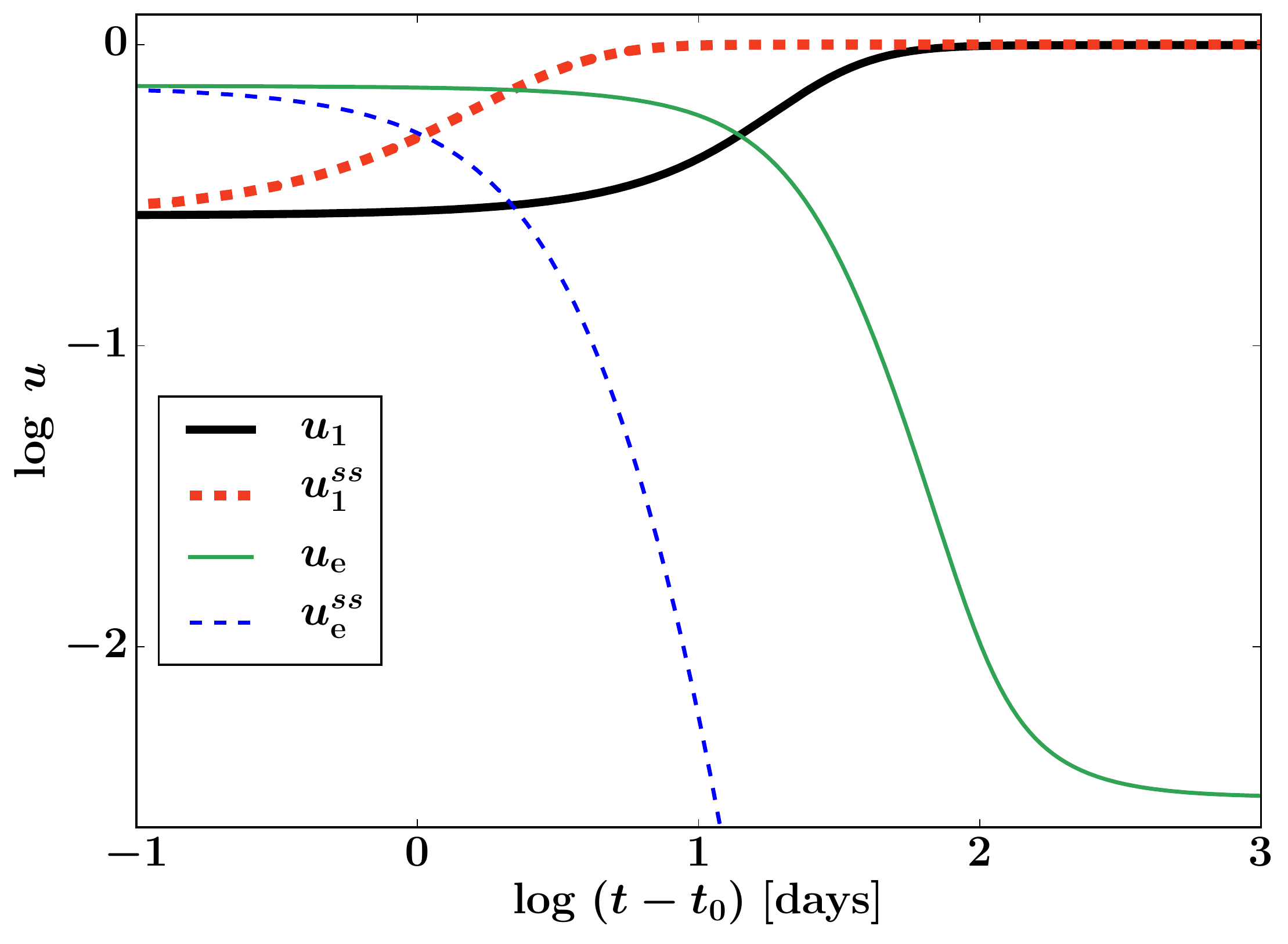}
    \caption {
    Number density functions
      $u_1(t)$, $u_1^{ss}(t)$, $\WI{u}{e}(t)$, and $\WI{u}{e}^{ss}(t)$
      depending on physical time ($t_0 = 20$~d).
    Calculations are carried out for the physical parameters typical for
      the near-photospheric layers
      of the SN~1999em
      (see Table~\ref{tab:values} column~2, Appendix~\ref{ap:values}).
    Initial conditions are ${u_1(t_0) = u_1^{ss}(t_0)}$ and
      $\WI{u}{e}(t_0) = \WI{u}{e}^{ss}(t_0)$.
    The normalized electron number density $\WI{u}{e}(t)$
      becomes constant $\WI{u}{e}(t{=}\infty) \approx 0.002$ at large time.
    }
    \label{fig:u_ss_ph}
  \end{minipage}
\end{figure*}

The ``freeze-out'' effect can be seen in
  Fig.~\ref{fig:u_ss_out} and~\ref{fig:u_ss_ph}.
The first figure presents a numerical calculation for the case of the outer, high-velocity layers
  far from the photosphere (Table~\ref{tab:values} column~1),
  while the second one is for physical parameters typical
  for the near-photospheric layers of SNe~IIP
  (Table~\ref{tab:values} column~2).
The calculations were carried out
  under the assumption that the system is initially in equilibrium.

The value of $\WI{u}{e}(\tau {=} \infty)$ may be small in some cases.
The deviation of true number densities from the equilibrium ones
  in Fig.~\ref{fig:u_ss_ph} at large time is not significant.
The solutions $u_1^{ss}$ and $u_1$ saturate to 1,
  while $\WI{u}{e}$, $\WI{u}{e}^{ss}$ are close to zero
  (which corresponds to almost full recombination).
However, there is a significant increase of deviation
  at the ${t-t_0\sim 10}$~d
  (Fig.~\ref{fig:u_ss_ph}).
For the outer, high-velocity layers far from the photosphere
  (Fig.~\ref{fig:u_ss_out})
  the deviation of number densities also increases
  over these days.

In the next section, we will investigate
  the factors that influence the evolution of the non-stationarity
  at early time, when the system is still close to equilibrium.

%

\section{The system at early time}
  \label{sec:early_time}

In Section~\ref{sec:large_time},
  the manifestation of violation of the steady-state approximation in kinetics was proved
  at large time.
However, its investigation by analytical methods is difficult
  because the Abel equation is not generally integrable by quadratures.
In this section, we are interested in
  deviation from the steady-state level populations at early time.
Numerical calculations for SN~1999em,
  which is a typical SN~IIP,
  show that at the initial time $t_0$ the deviation of real number densities
  from the equilibrium ones is negligible
  \citep{PotashovBlinnikovUtrobin2017a}.
In contrast to Section~\ref{sec:large_time}, where the qualitative picture
  was sufficient to solve the problem at large time,
  here more quantitative estimates are required.
Hence, the problem will be solved semi-analytically
  using typical values of the coefficients of the system
  (see Appendix~\ref{ap:values}).

Let us rewrite the system (\ref{eq:kinetics:td2:u1}, \ref{eq:kinetics:td2:ue}) in vector notation.
\begin{equation*}
  \dot{\mathbf{u}} = \mathbf{f}(\tau,\mathbf{u}),
\end{equation*}
  where $\mathbf{u} {=} {(u_1,\WI{u}{e})^\mathrm{T}}$ is the vector of normalized populations
  and $\mathrm{T}$ denotes the transpose of a row-vector into a column-vector.
We construct a system of equations in deviations
  $\mathbf{x}^{ss} {=} {(x_1^{ss},x_e^{ss})^{\mathrm{T}}}$
  of the true normalized populations $\mathbf{u}$
  from the steady-state solution of the system
  (\ref{eq:kinetics:ss:u1}, \ref{eq:kinetics:ss:ue})
  $\mathbf{u}^{ss} {=} {(u_1^{ss},\WI{u}{e}^{ss})^\mathrm{T}}$,
  that is, $\mathbf{x}^{ss} \equiv \mathbf{u} - \mathbf{u}^{ss}$.
Therefore, in general, one can write
\begin{equation*}
  \frac{d}{d\tau}(\mathbf{x}^{ss}+\mathbf{u}^{ss}) = \mathbf{f}(\tau,\mathbf{x}^{ss}+\mathbf{u}^{ss}),
\end{equation*}
and hence
\begin{equation}\label{eq:kinetics:ss:x}
  \dot{\mathbf{x}}^{ss} = \mathbf{f}(\tau,\mathbf{x}^{ss}+\mathbf{u}^{ss}) - \dot{\mathbf{u}}^{ss}.
\end{equation}
Because by definition ${\mathbf{f}(\tau,0+\mathbf{u}^{ss}) = 0}$,
  the value ${\mathbf{x}^{ss} = 0}$ is an equilibrium point.
Therefore, one can write
\begin{equation}\label{eq:kinetics:ss:x:lin}
  \dot{\mathbf{x}}^{ss} = \mathbf{J}(\tau)\,\mathbf{x}^{ss} + \mathbf{h}(\tau,\mathbf{x}^{ss}) - \dot{\mathbf{u}}^{ss},
\end{equation}
  where
  \[
  {\mathbf{J}(\tau) = \left.\frac{{\partial (f_1,\WI{f}{e})}}{{\partial (u_1^{ss},\WI{u}{e}^{ss})}}\right|_{u_i=u_i^{ss}}}
  \]
  is a Jacobi matrix and
  $\mathbf{h}(\tau,\mathbf{x}^{ss})$ represents higher-order terms.
\begin{equation}\label{eq:jacobian}
  \setlength\arraycolsep{2pt}
  \mathbf{J}(\tau) =
  \begin{pmatrix}
    -Q - \frac{A}{u_1^{ss}} \frac{(1-\WI{u}{e}^{ss})}{u_1^{ss}} \tau^2&
    -Q - \frac{A}{u_1^{ss}} \tau^2\\
    -\frac{P}{\tau^{s_2}}&
    -\frac{P}{\tau^{s_2}}-2 \WI{u}{e}^{ss} \frac{R}{\tau^3}\\
  \end{pmatrix}.
\end{equation}
Detailed analysis of $\mathbf{h}$
  shows that these terms vanish rapidly relative to ${\|\mathbf{x}\|}$
  as the origin is approached, and the relationship
  \[
  {\lim\limits_{\|\mathbf{x}\| \to 0} \sup_{\tau\,\geqslant\,1}
    \frac{\|\mathbf{h}(\tau,\mathbf{x})\|}{\|\mathbf{x}\|} = 0}
  \]
  holds.
Then the first-order approximation is accurate and defined as
  the linearization of the system
  \citep[see, for example,][p.~210]{Vidyasagar2002book}.
The final Cauchy problem can thus be formulated as
\begin{equation}\label{eq:kinetics:ss:x:lin2}
  \dot{\mathbf{x}}^{ss} = \mathbf{J}(\tau)\,\mathbf{x}^{ss}-\dot{\mathbf{u}}^{ss},
  \qquad
  \mathbf{x}^{ss}(1) = 0,
\end{equation}
  where we assume the initial conditions to be zero.

The linear time-varying system (\ref{eq:kinetics:ss:x:lin2})
  describes the behaviour of the system
  (\ref{eq:kinetics:ss:x}) in a small segment of initial time
  while deviations $\mathbf{x}^{ss}$ are small.
We will be able to understand how the deviations of $\mathbf{x}^{ss}$
  from the steady-state solution grow by analysing the linearized case.

The general solution of (\ref{eq:kinetics:ss:x:lin2}) in vector form is written as
  \multicitep{Demidovich1967book, p.~77; Kuijstermans2003thesis, p.~41}
\begin{equation}\label{eq:kinetics:ss:x:lin2:sol}
  \mathbf{x}^{ss}(\tau) =
    -\int\limits_{1}^{\tau}\mathbf{K}(\tau,\tau')\,\dot{\mathbf{u}}^{ss}(\tau') d\tau',
\end{equation}
  where ${\mathbf{K}(\tau, \tau') = \mathbf{X}(\tau)\mathbf{X}(\tau')^{-1}}$
  is a normalized fundamental matrix at point $\tau'$,
  and $\mathbf{X}(\tau)$ is fundamental matrix.
$\mathbf{K}(\tau, \tau')$ is also called a matriciant or Cauchy matrix.
Our goal is to obtain an expression for $\mathbf{K}(\tau, \tau')$.
To find the fundamental matrix $\mathbf{X}(\tau)$,
  it is convenient to introduce the concept of
  dynamic eigenvalues and eigenvectors
  \multicitep{Wu1980; NeerhoffVanderKloet2001; Kuijstermans2003thesis, p.~47}.
If there exists a scalar function $\lambda(\tau)$
  and a non-zero differentiable vector function $\mathbf{v}(\tau)$,
  so that they satisfy the following condition
\begin{equation}\label{eq:eigen_dynamic}
  [\mathbf{J}(\tau)-\lambda(\tau)\mathbf{I}]\,\mathbf{v}(\tau) = \dot{\mathbf{v}}(\tau),
\end{equation}
  where $\mathbf{I}$ is an identity matrix,
  then $\lambda(\tau)$ is called a \emph{dynamic eigenvalue}
  of matrix $\mathbf{J}(\tau)$ associated with
  a \emph{dynamic eigenvector} $\mathbf{v}(\tau)$.
Quasi-static $\lambda^{qs}(\tau)$ and $\mathbf{v}^{qs}(\tau)$
  are the classical eigenvalues and eigenvectors
  obtained by solving equation
  (\ref{eq:eigen_dynamic}) with the right-hand side equal to zero.
Similar to the classical algebraic case,
  the number of distinct dynamic eigenvalues $\lambda_i(\tau)$
  of the matrix is equal to or less than the size of the matrix.
The number of all dynamic eigenvalues of $\mathbf{J}$
  is equal to the size of the matrix.

Furthermore, we change the variables
  using the transformation
  ${\mathbf{x}^{ss} = \mathbf{L}(\tau) \mathbf{y}}$,
  where $\mathbf{y} = \mathbf{y}(\tau)$
  is a vector of new unknown variables.
The $\mathbf{J}$ matrix is converted to a new matrix,
\begin{equation*}
  \mathbf{B} = \mathbf{L}^{-1}\mathbf{J}\mathbf{L} - \mathbf{L}^{-1}\dot{\mathbf{L}}.
\end{equation*}
In the classical algebraic case
  for constant $\mathbf{J}$ and $\mathbf{L}$,
  $\mathbf{B}$ is Jordan matrix
  whenever $\mathbf{J}$ has repeated eigenvalues.
In the case with time-varying matrices, one can choose
  an algebraic transformation $\mathbf{L}$
  so that $\mathbf{B}$ transforms
  into a diagonal $ \mathbf{\Lambda}$ matrix
  \citep[Theorem~2]{Wu1980}.
This transformation belongs to the class of Lyapunov transformations
  \citep[p.~133]{Kuijstermans2003thesis}.
Because the coordinate transformation $\mathbf{L}$ preserves
  the dynamic eigenvalues
  \citep{Wu1980, NeerhoffVanderKloet2001},
  ${\mathbf{\Lambda(\tau)} = \mathrm{diag}[\lambda_i(\tau)]}$.
The final expression for the fundamental matrix
  $\mathbf{X}(\tau)$ is
\begin{equation}\label{eq:matriciant_dynamic}
  \mathbf{X}(\tau) = \mathbf{L}(\tau)\,\mathrm{diag}[e^{\gamma_i(\tau)}],
\end{equation}
  where
\begin{equation}\label{eq:gamma}
  \gamma_i(\tau) \equiv \int\limits_{1}^{\tau} \lambda_i(\zeta') d \zeta'.
\end{equation}
Thus, the problem of finding solutions of (\ref{eq:kinetics:ss:x:lin2})
  is reduced to the search of its eigenvalues $\lambda_i(\tau)$
  and the parameters for a Lyapunov transformation
  $\mathbf{L}$ that obtains the diagonal matrix ${\mathbf{\Lambda}}$.

One way to find $\mathbf{L}(\tau)$ has been proposed by
  \citet{Wu1980, VanderKloetNeerhoff2000};
  \citet[p.~137]{Kuijstermans2003thesis}.
This method is based on an iterative algorithm
\begin{equation*}
  \mathbf{\bar{\Lambda}}_j = \mathbf{Q}_j^{-1}(\mathbf{\bar{\Lambda}}_{j-1}
    -\mathbf{Q}_{j-1}^{-1}\dot{\mathbf{Q}}_{j-1})\mathbf{Q}_j
    \quad
    (j = 1,2,\dots),
\end{equation*}
  with the conditions
\begin{equation*}
  \mathbf{\bar{\Lambda}}_0 = \mathbf{J},
    \quad
    \mathbf{Q}_0 = \mathbf{I},
\end{equation*}
  where $\mathbf{I}$ is an identity matrix,
  $\mathbf{\bar{\Lambda}}_j$ is a diagonal matrix,
  and $\mathbf{Q}_j$ is a transformation matrix that consists
  of quasi-static eigenvectors of the matrix
  ${\mathbf{\bar{\Lambda}}_{j-1}-\mathbf{Q}_{j-1}^{-1}\dot{\mathbf{Q}}_{j-1}}$,
  calculated at every moment of time.
Thus, at each step of the iteration,
  the matrix from the previous step is diagonalized
  taking into account the error
  ${\mathbf{Q}_{j-1}^{-1}\dot{\mathbf{Q}}_{j-1}}$.
\citet{VanderKloetNeerhoff2000} have proved
  that the iterative process converges:
\begin{gather*}
  \lim_{j \to \infty} \mathbf{\bar{\Lambda}}_j(\tau) = \mathbf{\Lambda}(\tau),
  \\
  \lim_{j \to \infty} \mathbf{Q}_1(\tau)\mathbf{Q}_2(\tau)\dots\mathbf{Q}_j(\tau) = \mathbf{L}(\tau).
\end{gather*}

Let us compare the norms of the diagonal matrix
  and the error at each step of the iteration.
For parameters typical for the problem
  that we are solving (see Appendix~\ref{ap:values}),
  the following inequality already holds for the first iterative step:
\begin{equation*}
  \lVert \mathbf{\bar{\Lambda}}_{1}(\tau) \rVert =
    \lVert \mathbf{Q}_{1}^{-1}(\tau)\mathbf{J}(\tau)\mathbf{Q}_{1}(\tau) \rVert
  \gg
  \lVert \mathbf{Q}_{1}^{-1}(\tau)\dot{\mathbf{Q}}_{1}(\tau) \rVert.
\end{equation*}
Indeed, for the initial moment, the norms can be estimated as
\begin{equation*}
  \lVert \mathbf{\bar{\Lambda}}_{1}(1) \rVert > P
  \gg 1 >
  \lVert \mathbf{Q}_{1}^{-1}(1)\dot{\mathbf{Q}}_{1}(1) \rVert,
\end{equation*}
  where $P$ is defined in (\ref{eq:abr2}).

The inequality is strengthened as
  ${\lVert \mathbf{\bar{\Lambda}}_{1}(\tau) \rVert \sim \mathcal{O}(\tau^2)}$
  with time ${\tau \to \infty}$ and
  ${\lVert \mathbf{Q}_{1}^{-1}(\tau)\dot{\mathbf{Q}}_{1}(\tau) \rVert \sim \mathcal{O}(1/\tau)}$.
Therefore, we can conclude
  that in our case the following approximation is perfectly suitable:
\begin{equation*}
  \mathbf{\Lambda}(\tau) \approx \mathrm{diag}[\lambda_i^{qs}(\tau)],
  \qquad
  \mathbf{L}(\tau) \approx \mathbf{V}_{qs}(\tau) = \mathbf{Q}_{1}(\tau),
\end{equation*}
  where $\mathbf{V}_{qs} {=} (\mathbf{v}_1^{qs}, \mathbf{v}_2^{qs})$
  is the matrix with quasi-static eigenvector columns
  $\mathbf{v}_1^{qs}$, $\mathbf{v}_2^{qs}$
  of the matrix $\mathbf{J}$
  corresponding to the quasi-static eigenvalues of the Jacobian matrix.
Therefore, expression (\ref{eq:matriciant_dynamic})
  for the fundamental matrix $\mathbf{X}$
  can be rewritten as
\begin{equation}\label{eq:matriciant_quasi_static}
  \mathbf{X}(\tau) = \mathbf{V}_{qs}(\tau)\,\mathrm{diag}[e^{\gamma^{qs}_i(\tau)}]
    =
    \left(
      \mathbf{v}_1^{qs}(\tau)\, e^{\gamma_1^{qs}(\tau)},
      \mathbf{v}_2^{qs}(\tau)\, e^{\gamma_2^{qs}(\tau)}
    \right),
\end{equation}
  where $\gamma_i^{qs}(\tau)$ is calculated from (\ref{eq:gamma})
  for $\lambda_i^{qs}$.
The normalized fundamental matrix looks like
\begin{equation}\label{eq:matrix_cauchy}
  \mathbf{K}(\tau, \tau')
  =
  \mathbf{V}_{qs}(\tau)\mathrm{diag}[\exp(\gamma_i^{qs}(\tau){-}\gamma_i^{qs}(\tau'))]\mathbf{V}_{qs}^{-1}(\tau').
\end{equation}

The quasi-static eigenvalues and eigenvectors
  for the two-dimensional matrix $\mathbf{J}$ can be written explicitly as
\begin{equation}\label{eq:eigen_values}
  \lambda_{1,2}^{qs}(\tau) =
    \frac{1}{2} \left(\mathrm{tr}(\mathbf{J}) \pm \sqrt{\mathrm{tr}(\mathbf{J})^2 - 4 \Delta}\right),
\end{equation}
\begin{equation}\label{eq:eigen_vectors}
  \mathbf{v}_{1,2}^{qs}(\tau) = \left(\frac{\lambda_{1,2}^{qs}(\tau) - \mathbf{J}[2,2]}{\mathbf{J}[2,1]},\,1\right)^{\mathrm{T}}.
\end{equation}
Here, $\mathrm{tr}(\mathbf{J})$ is the trace of the Jacobi matrix
  and $\Delta$ is its determinant.
It should be noted that
  $\mathrm{tr}(\mathbf{J}) = \lambda_{1}^{qs}+\lambda_{2}^{qs}$
  and $\Delta = \lambda_{1}^{qs} \lambda_{2}^{qs}$.

From (\ref{eq:kinetics:ss:ue:power}), it follows that
  the function $\WI{u}{e}^{ss}/\tau^3$ decreases as
  ${\sim \tau^{-(s_1 + s_2 + 3)/2}}$,
  which is steeper than $\tau^{-s_2}$.
Because of $P \gg R$ (see Appendix~\ref{ap:values}),
  ${P/\tau^{s_2} \gg 2 R\, \WI{u}{e}^{ss}/\tau^3}$
  for any time ${\tau \geqslant 1}$.
Therefore, $\mathbf{J}[2,2] \simeq \mathbf{J}[2,1]$
  and
\begin{equation}\label{eq:eigen_vectors2}
  \mathbf{v}_{1,2}^{qs}(\tau) \simeq \left(-1 - \frac{\lambda_{1,2}^{qs}(\tau)}{P}\tau^{s_2},\,1\right)^{\mathrm{T}}
\end{equation}
  according to (\ref{eq:jacobian}).

Consider now the first eigenvalue of the matrix $\mathbf{J}$
  defined in (\ref{eq:eigen_values}).
  It is equal to
\begin{equation}\label{eq:lambda1}
  \lambda_{1}^{qs}(\tau)
    =
    -\frac{\WI{u}{e}^{ss} \frac{R}{\tau^3} \left(2\,Q\,u_1^{ss} + \left(\frac{1}{u_1^{ss}}+1\right) A\tau^2\right)}
    {A\tau^2+u_1^{ss} P/\tau^{s_2}},
\end{equation}
  which takes into account (\ref{eq:jacobian}).
The expression~(\ref{eq:lambda1}) implies that $\lambda_{1}^{qs}(\tau)$
  is a negative time function.
One can ensure that initially
  $|\lambda_{1}^{qs}(1)| \ll P$
  (see Appendix~\ref{ap:values}).
As time increases, the module of the function $|\lambda_{1}^{qs}(\tau)|$
  decreases monotonically faster than
  $\WI{u}{e}^{ss} R/\tau^3 \sim \tau^{-(s_1 + s_2 + 3)/2}$,
  which is steeper than $\tau^{-s_2}$.
This means that $|\lambda_{1}^{qs}(\tau)| \ll P/\tau^{s_2}$
  for any ${\tau \geqslant 1}$.
Thus, the first eigenvector is simplified to
\begin{equation}
  \mathbf{v}_{1}^{qs}(\tau) \simeq (-1,\,1)^{\mathrm{T}}.
\end{equation}

Consider now the behaviour of the function $\mathrm{tr}(\mathbf{J}(\tau))$.
This is a negative function of $\tau$.
For its module, the inequality $|\lambda_{1}^{qs}(1)| \ll P < |\mathrm{tr}(\mathbf{J}(1))|$ holds.
Because $P \gg A \gg R$
  (see Appendix~\ref{ap:values}),
  the module $|\mathrm{tr}(\mathbf{J}(\tau))|$ initially
  decreases as $P/\tau^{s_2}$ with time
  not steeper than $|\lambda_{1}^{qs}(\tau)|$
  and then increases as $A \tau^2$.
Hence, $|\lambda_{1}^{qs}(\tau)| \ll |\mathrm{tr}(\mathbf{J}(\tau))|$
  for any ${\tau \geqslant 1}$.
And finally, from
\begin{equation}\label{eq:lambda2}
  \lambda_{2}^{qs}(\tau) = \mathrm{tr}(\mathbf{J}(\tau)) - \lambda_{1}^{qs}(\tau)
\end{equation}
  it follows that $|\lambda_{1}^{qs}(\tau)| \ll |\lambda_{2}^{qs}(\tau)|$
  and $\lambda_{2}^{qs}(\tau) \simeq \mathrm{tr}(\mathbf{J}(\tau))$
  for any ${\tau \geqslant 1}$.
Thus, the second eigenvector can be rewritten as
\begin{equation}
  \mathbf{v}_{2}^{qs}(\tau) \simeq \left(-1-\frac{\mathrm{tr}(\mathbf{J}(\tau))}{P}\tau^{s_2},\,1\right)^{\mathrm{T}}.
\end{equation}

For small time,
  as long as $|\mathrm{tr}(\mathbf{J}(\tau))|$
  is dominated by the photoionization term $P/\tau^{s_2}$, we have
\begin{equation*}
  \int\limits_{\tau'}^{\tau} \lambda_2^{qs}(\zeta) d \zeta
    \simeq
    -\int\limits_{\tau'}^{\tau} \frac{P}{\zeta^{s_2}} d \zeta
    \simeq
    -\frac{P}{\tau^{s_2}}\,(\tau{-}\tau') + \mathcal{O}(\tau{-}\tau')^2,
\end{equation*}
but at large time, when the spontaneous emission $A\,\tau^{2}$ dominates, we can write
\begin{equation*}
  \int\limits_{\tau'}^{\tau} \lambda_2^{qs}(\zeta) d \zeta
    \simeq
    -\int\limits_{\tau'}^{\tau} A \zeta^{2} d \tau
    \simeq
    -A \tau^{2}\,(\tau{-}\tau') + \mathcal{O}(\tau{-}\tau')^2.
\end{equation*}
Because
\begin{equation*}
  \mathrm{min}(|\lambda_{2}^{qs}(\tau)|) > A\left(\frac{P}{A}\right)^{\frac{2}{2+s_2}} \gg 1,
\end{equation*}
  one has that
  ${\exp(\gamma_2^{qs}(\tau)-\gamma_2^{qs}(\tau'))}$
  is an exponentially decreasing function of
  ${(\tau-\tau')}$, for any ${\tau \geqslant 1}$.
One can estimate the normalized fundamental matrix
  by summarizing all the above observations.
From (\ref{eq:matrix_cauchy}),
  it follows that when $\tau'$ and $\tau$ are nearly close,
\begin{equation}\label{eq:t-t1}
  (\tau-\tau') \lesssim \mathrm{min}(|\lambda_{1}^{qs}(\tau)|)^{-1} \ll 1,
\end{equation}
  $\mathbf{K}(\tau, \tau') \sim \mathbf{I}$.
In the opposite case, we assume that
  ${\exp(\gamma_2^{qs}(\tau)-\gamma_2^{qs}(\tau')) \simeq 0}$
and find
\begin{equation*}
  \mathbf{K}(\tau, \tau') \simeq
    \frac{
      \exp(\gamma_1^{qs}(\tau)-\gamma_1^{qs}(\tau'))
      }
      {1+\mathbf{v}_{2}^{qs}[1](\tau')}
    \setlength\arraycolsep{2pt}
    \begin{pmatrix}
      \phantom{-}1 &
      -\mathbf{v}_{2}^{qs}[1](\tau') \\
      -1 &
      \phantom{-}\mathbf{v}_{2}^{qs}[1](\tau') \\
    \end{pmatrix}.
\end{equation*}

Finally, we consider the integrand function in (\ref{eq:kinetics:ss:x:lin2:sol}).
If (\ref{eq:t-t1}) is true
  then $\mathbf{K}(\tau,\tau')\,\dot{\mathbf{u}}^{ss}(\tau') \sim \dot{\mathbf{u}}^{ss}(\tau)$,
  and the contribution of this term can be neglected.
When (\ref{eq:t-t1}) is not satisfied, the estimate can be written as
\begin{equation*}
  \mathbf{K}(\tau,\tau')\,\dot{\mathbf{u}}^{ss}(\tau') \simeq
    \exp(\gamma_1^{qs}(\tau)-\gamma_1^{qs}(\tau'))\,
    M(\tau')
    \begin{pmatrix}
      \phantom{-}1 \\
      -1 \\
    \end{pmatrix},
\end{equation*}
  where
\begin{equation}\label{eq:M}
  \begin{multlined}
    M(\tau) =
      \frac{\dot{u}_1^{ss}(\tau) - \WI{\dot{u}}{e}^{ss}(\tau)\,\mathbf{v}_{2}^{qs}[1](\tau)}
      {1+\mathbf{v}_{2}^{qs}[1](\tau)} =
      \\
      {=}
      \dot{u}_1^{ss}(\tau) {-}
      \left(\!
        1{+}
        \frac{P}{\mathrm{tr}(\mathbf{J}(\tau))\,\tau^{s_2}}
      \!\right)
      (\dot{u}_1^{ss}(\tau) {+} \WI{\dot{u}}{e}^{ss}(\tau)).
  \end{multlined}
\end{equation}
The analysis of the behaviour of the function $\mathrm{tr}(\mathbf{J}(\tau))$
  presented above implies that
  \[
  \left(1+\frac{P}{\mathrm{tr}(\mathbf{J}(\tau))\,\tau^{s_2}}\right) \lesssim 2.
  \]
And because ${|(\dot{u}_1^{ss}(\tau) + \WI{\dot{u}}{e}^{ss}(\tau))| \ll |\dot{u}_1^{ss}(\tau)|}$
  for any ${\tau \geqslant 1}$,
  we obtain the final answer
\begin{equation}\label{eq:kinetics:ss:x:lin2:sol2}
  \mathbf{x}^{ss}(\tau) {\simeq}
    \!\!\int\limits_{1}^{\tau}\!\!
    \exp(\gamma_1^{qs}(\tau){-}\gamma_1^{qs}(\tau'))
    \,\dot{u}_1^{ss}(\tau') d\tau'
    \!
    \begin{pmatrix}
      -1 \\
      \phantom{-}1 \\
    \end{pmatrix}.
\end{equation}
The solution (\ref{eq:kinetics:ss:x:lin2:sol2}) includes only the larger
  of the two eigenvalues $\lambda_{1}^{qs}(\tau)$,
  which is the smaller in absolute value
  because they are negative.
The smaller the $|\lambda_{1}^{qs}(\tau)|$ module,
  the faster the growth of deviation from the steady state, and the greater
  the role of time dependence.
Thus, the value of this eigenvalue is a key parameter
  influencing the time-dependent effect.

\begin{figure*}
  \begin{minipage}{0.47\textwidth}\centering
    \includegraphics[width=\textwidth]{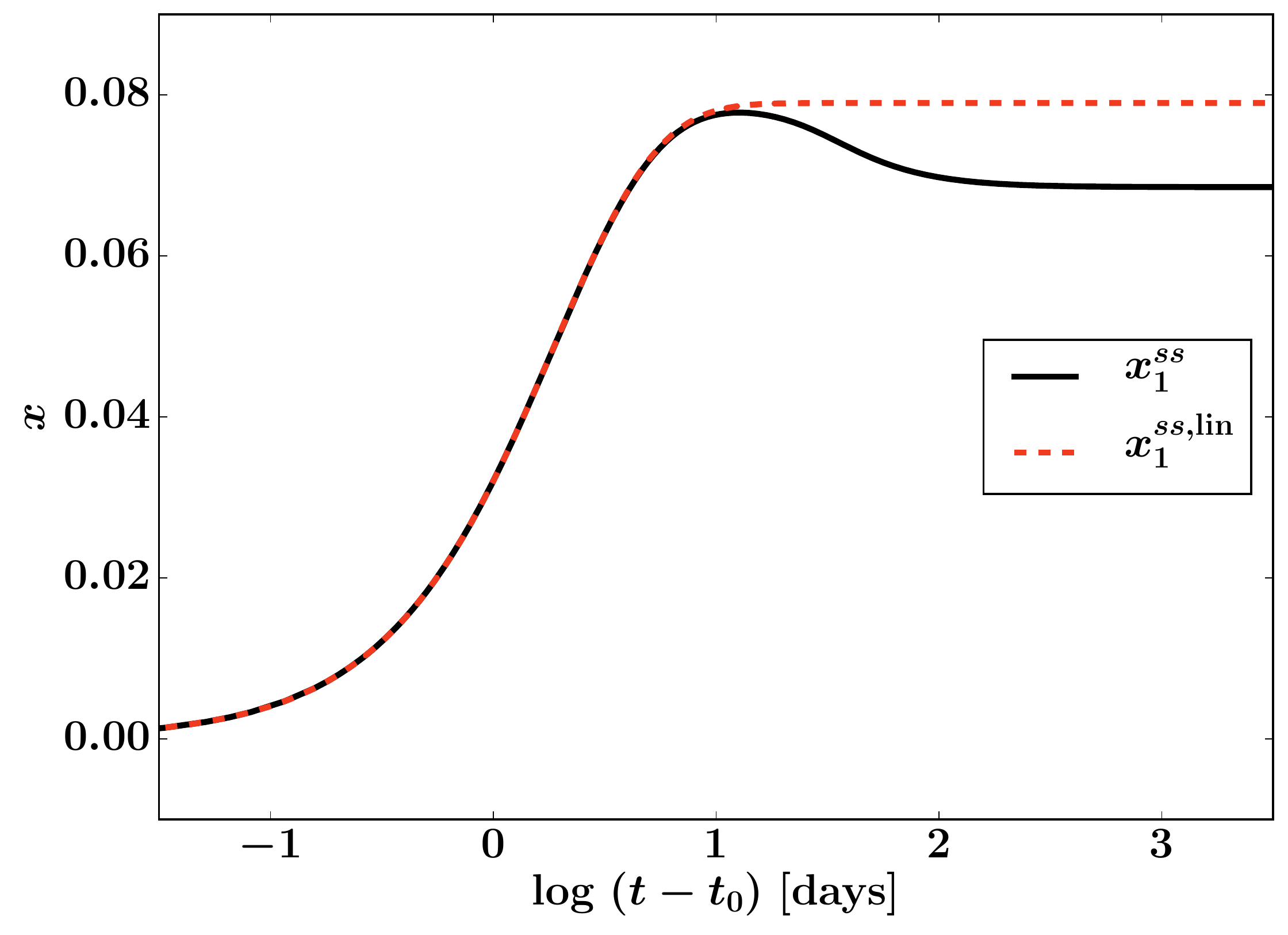} \\ a)
  \end{minipage}\hfill
  \begin{minipage}{0.47\textwidth}\centering
    \includegraphics[width=\textwidth]{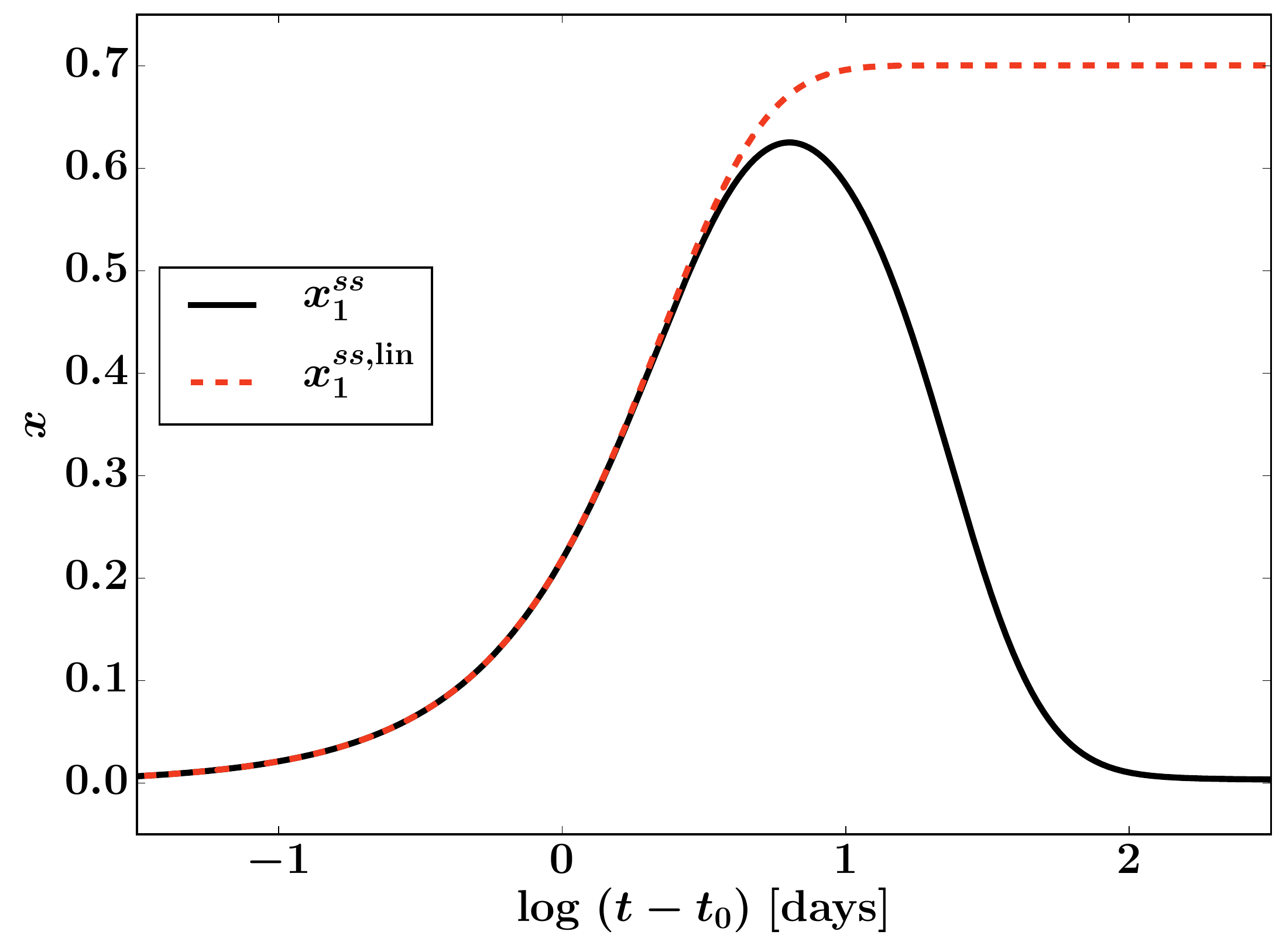} \\ b)
  \end{minipage}
  \caption {
  The functions $x_1^{ss}(t)$, $x_1^{ss,\mathrm{lin}}(t)$
    calculated under the same conditions
    as for Fig.~\ref{fig:u_ss_out}(a)
    and for Fig.~\ref{fig:u_ss_ph}(b)
    on physical time (${t_0 {=} 20}$~d).
  It is clear that the $x_1^{ss,\mathrm{lin}}(t)$
    correctly reproduces the evolution of $x_1^{ss}(t)$
    only at early time (for ${t-t_0\sim 10}$~d)
    in both cases.
  }
  \label{fig:x1_ss}
\end{figure*}

The linearization (\ref{eq:kinetics:ss:x:lin2})
  of the system (\ref{eq:kinetics:ss:x})
  is applicable only when the deviations
  ${\mathbf{x}^{ss}}$ are small.
However, we have previously seen that the deviations can be significant.
Fig.~\ref{fig:x1_ss}(a) presents the solution $x_1^{ss}(t)$
  of the system (\ref{eq:kinetics:ss:x:lin2})
  and its linearization $x_1^{ss,\mathrm{lin}}(t)$
  derived from (\ref{eq:kinetics:ss:x:lin2:sol2})
  for physical time for the case of outer, high-velocity layers
  far from the photosphere
  (Table~\ref{tab:values} column~1).
  while Fig.~\ref{fig:x1_ss}(b) presents
  solutions of the same systems
  for the case of near-photospheric layers
  (Table~\ref{tab:values} column~2).
The strong nonlinearity of the solutions $x_1^{ss}(t)$ is clearly seen
  when the normalized number densities $u_1^{ss}$ and $u_1$
  are saturated to 1, which corresponds to almost full recombination
  (see Fig.~\ref{fig:u_ss_out},\ref{fig:u_ss_ph}).
Thus, the solution (\ref{eq:kinetics:ss:x:lin2:sol2}) describes
  the evolution of the system only at \emph{early} time.

However, in the next section,
  we show that the final expression for $|\lambda_{1}^{qs}(\tau)|$
  determines whether the kinetic system is in equilibrium
  at \emph{any} moment of time.

\section{Frozen system}
  \label{sec:frozen_system}

The question remains wether it is possible to specify a simple way
  to check the importance of the time-dependence effect under given conditions.
The solution of the system with ``frozen'' coefficients would help us to find the answer
  \multicitep{Khalil2002book, p.~365; Vidyasagar2002book, p.~248}.

Let us rewrite the system of equations
  (\ref{eq:kinetics:td2:u1}, \ref{eq:kinetics:td2:ue})
  where all the variable coefficients are fixed at some moment $\tau_1$.
In fact, a non-autonomous system have transformed into an autonomous one.
\begin{equation}\label{eq:kinetics:ss2:u1}
  \dot{u}_{1}^{a} = (1-u_1^{a}-\WI{u}{e}^{a})\;\left(Q + \frac{A}{u_1^{a}}\tau_1^2\right)
    -\frac{B}{\tau_1^{s_1-2}},
\end{equation}
\begin{equation}\label{eq:kinetics:ss2:ue}
  \WI{\dot{u}}{e}^{a} = (1-u_1^{a}-\WI{u}{e}^{a})\frac{P}{\tau_1^{s_2}}
    - (\WI{u}{e}^{a})^2\frac{R}{\tau_1^3}.
\end{equation}
Unknown ${u_1^{a}(\tau_a)}$, ${\WI{u}{e}^{a}(\tau_a)}$
  are the functions of a new time variable $\tau_a$.
Obviously, the values of expressions $u_1^{ss}(\tau_1)$, $\WI{u}{e}^{ss}(\tau_1)$,
  are the solutions of the system
  (\ref{eq:kinetics:ss:u1}, \ref{eq:kinetics:ss:ue}).
The right-hand side of
  (\ref{eq:kinetics:ss2:u1}, \ref{eq:kinetics:ss2:ue})
  vanishes for those values.

Let us prove that for any solution of the system
  (\ref{eq:kinetics:ss2:u1}, \ref{eq:kinetics:ss2:ue})
  for $\tau_a \to \infty$
  under initial conditions $0 < u_{1,e}^{a}(\tau_a=1) \leqslant 1$
  one obtains $u_{1,e}^{a}(\tau_a) \to u_{1,e}^{ss}(\tau_1)$.

The change of variables
  $x_1(\tau_a) = u_1^{a}(\tau_a) - u_1^{ss}(\tau_1)$
  and $\WI{x}{e}(\tau_a) = \WI{u}{e}^{a}(\tau_a) - \WI{u}{e}^{ss}(\tau_1)$
  transforms the original system to the reduced one
  \multicitep{Demidovich1967book, p.~234; Khalil2002book, p.~147}:
\begin{equation}\label{eq:kinetics:ss2:x1}
  \dot{x}_1 = -(x_1{+}\WI{x}{e})\,Q
    -
    \frac{(1-\WI{u}{e}^{ss}(\tau_1))\,x_1+u_1^{ss}(\tau_1)\,\WI{x}{e}}
    {u_1^{ss}(\tau_1)(u_1^{ss}(\tau_1)+x_1)}
    A\tau_1^2,
\end{equation}
\begin{equation}\label{eq:kinetics:ss2:xe}
  \WI{\dot{x}}{e} = -(x_1{+}\WI{x}{e})\frac{P}{\tau_1^{s_2}}
    -\WI{x}{e}\;(2\,\WI{u}{e}^{ss}(\tau_1){+}\WI{x}{e})\frac{R}{\tau_1^3}.
\end{equation}
The Jacobi matrix of this system
  coincides with the matrix (\ref{eq:jacobian})
  taken at time $\tau_1$: ${\mathbf{J} = \mathbf{J}(\tau_1)}$.
From the analysis carried out in Section~\ref{sec:early_time},
  it follows that the two eigenvalues $\lambda_1$, $\lambda_2$
  of the Jacobi matrix $\mathbf{J}$ are negative and different.
Consequently, by Lyapunov's indirect method
  \multicitep{Fedoruk1985book, p.~289; Khalil2002book, p.~133}, it appears that
  the system (\ref{eq:kinetics:ss2:x1}, \ref{eq:kinetics:ss2:xe})
  is exponentially stable,
  and its solution is written as
  \multicitep{Fedoruk1985book, p.~61; Khalil2002book, p.~37}
\begin{equation}\label{eq:kinetics:ss2:x:sol}
  \mathbf{x}(\tau_a) = (x_1, \WI{x}{e})^\mathrm{T} = C_1 e^{\lambda_1 \tau_a} \mathbf{v}_1 + C_2 e^{\lambda_2 \tau_a} \mathbf{v}_2,
\end{equation}
  where $\mathbf{v}_i$ are the eigenvectors of the Jacobi matrix $\mathbf{J}$
  and $C_i$ are arbitrary constants.
Obviously, $\lim_{\tau \to \infty} \mathbf{x}(\tau_a) = 0$,
  which shows the convergence property of the system
  (\ref{eq:kinetics:ss2:u1}, \ref{eq:kinetics:ss2:ue})
  \citep[p.~281]{Demidovich1967book}.
Thus, the search for the limit of solutions of
  (\ref{eq:kinetics:ss2:u1}, \ref{eq:kinetics:ss2:ue})
  provides another way to find
  the algebraic approximation of $u_1^{ss}$, $u_e^{ss}$.

\begin{figure*}
  \begin{minipage}{0.47\textwidth}\centering
    \includegraphics[width=\textwidth]{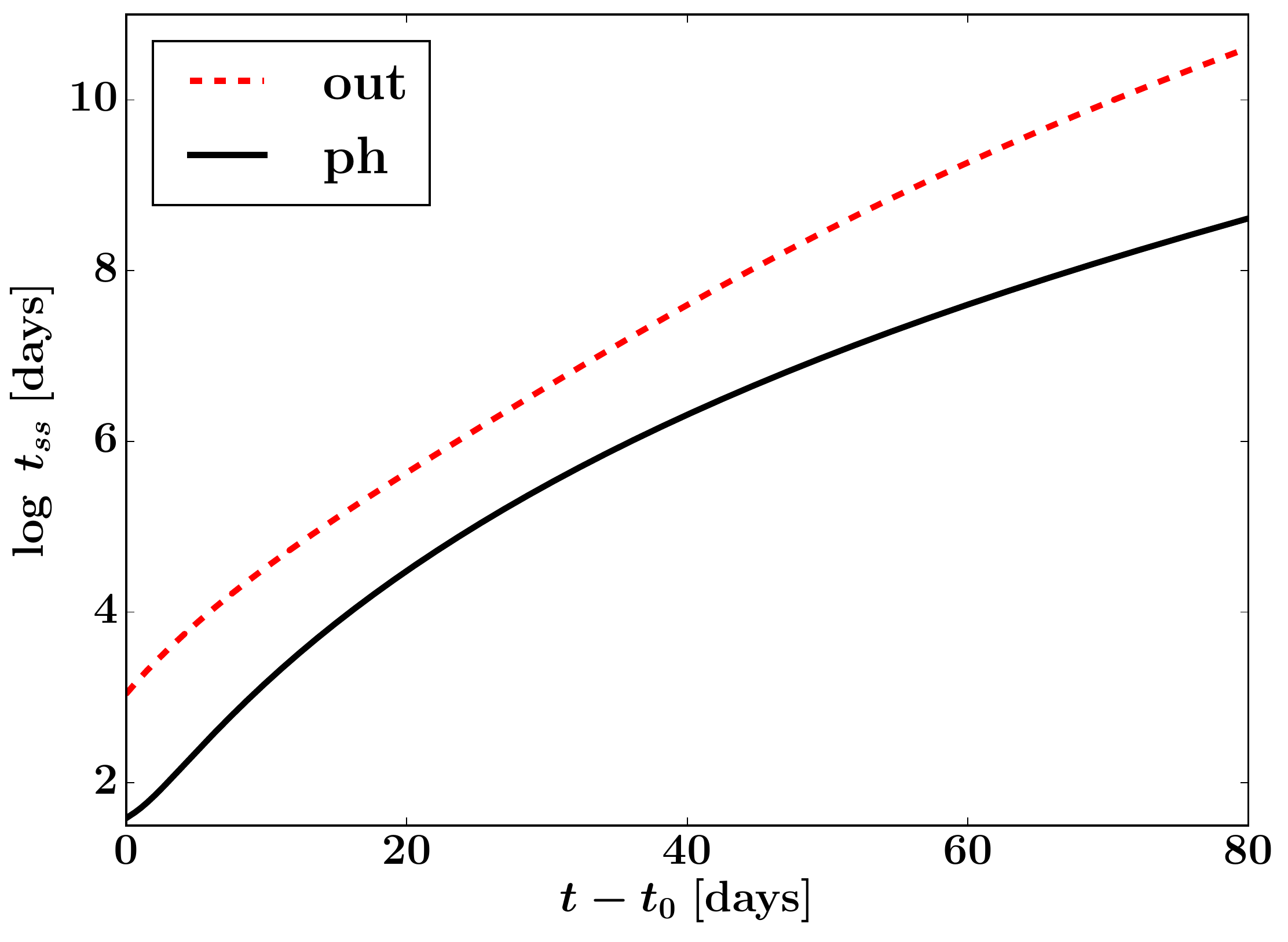}
    \caption {
    The relaxation time ${t_{ss} = \tau_{ss} t_0}$,
      in the case of large optical depth
      for the outer, high-velocity layers (label ``out'')
      and the near-photospheric layers (label ``ph'')
      with respect to physical time
      with ${t_0 {=} 20}$~d
      (see Appendix~\ref{ap:values}).
    }
    \label{fig:tss}
  \end{minipage}\hfill
  \begin{minipage}{0.47\textwidth}\centering
    \includegraphics[width=\textwidth]{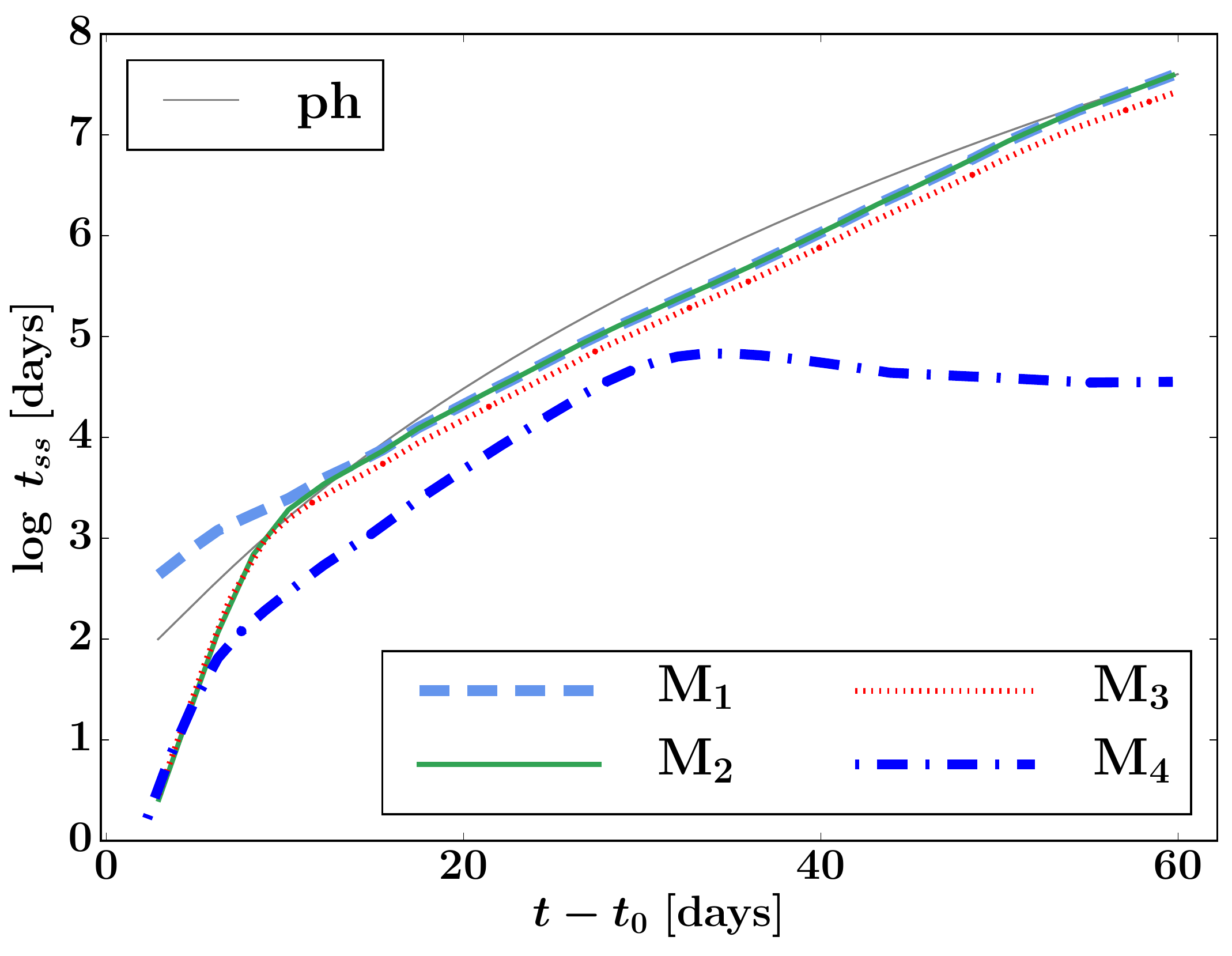}
    \caption {
    Illustration of the numerical calculation of relaxation times
      ${t_{ss} = \tau_{ss} t_0}$
      for the four models listed in Table~\ref{tab:systems}.
    The thin solid line fully coincides
      with the ``ph''--curve in Fig.~\ref{fig:tss}.
    The abscissa is the physical time, where ${t_0 {=} 20}$~d.
    }
    \label{fig:tss_full}
  \end{minipage}
\end{figure*}

The next question that arises is how long a period
  $\tau_{ss}$ is needed for the solution $u_1^{a}(\tau_a)$ to obtain
  the stationary value $u_1^{ss}$.
If this period is significantly longer than the time-scale
  of change of the envelope's physical parameters,
  the system does not have time to relax to equilibrium.
From (\ref{eq:kinetics:ss2:x:sol}), it follows that
  the period of equilibration
  is determined by the smaller of the two modules of the eigenvalues
  ${\tau_{ss}^{-1} = \mathrm{min}(|\lambda_1(\tau_1)|, |\lambda_2(\tau_1)|)}$.
From Section~\ref{sec:early_time}, we know that
  ${|\lambda_{1}(\tau_1)| \ll |\lambda_{2}(\tau_1)|}$,
  and thus ${\tau_{ss}^{-1} = |\lambda_1(\tau_1)|}$ holds.
The relaxation time $\tau_{ss}$ for any $\tau_1$ from (\ref{eq:lambda1}) is
\begin{equation}\label{eq:tau_ss}
  \tau_{ss}(\tau)
    =
    \frac{A\tau^2+u_1^{ss} \frac{P}{\tau^{s_2}}}
    {\WI{u}{e}^{ss} \frac{R}{\tau^3} \left(2\,Q\,u_1^{ss} + \left(\frac{1}{u_1^{ss}}+1\right) A\tau^2\right)}.
\end{equation}
A simple criterion for checking the equilibrium of the system
  during the photospheric phase at any time arises from a comparison
  of the value of expression (\ref{eq:tau_ss})
  with the typical time-scale of change of SN envelope's physical parameters.
The calculations show (Fig.~\ref{fig:tss})
  that in the case of an optically thick medium,
  for the outer, high-velocity layers (``out'' label)
  and for the near-photospheric layers (``ph'' label),
  the time ${t_{ss} = \tau_{ss} t_0}$ substantially exceeds
  the duration of the photospheric phase itself.
This means that the system is always in the non-equilibrium state
  for the whole envelope.
The time-dependence effect is more significant for the outer layers.

The value of $\tau_{ss}$
  increases with time (Fig.~\ref{fig:tss}),
  so at ${\tau \to \infty}$ the time-dependent effect
  will always be important,
  which has already been proved in Section~\ref{sec:large_time}.
However, we have seen (Fig.~\ref{fig:u_ss_ph}) that,
  for example, in the near-photospheric layers
  the deviations of the number densities
  from the steady-state values can decrease
  when the atoms are almost fully recombined.
It can be concluded that the normalized populations
  of the equilibrium and non-equilibrium systems
  are close only when these populations are saturated.

Let us determine the physical quantities
that  affect the relaxation time more than others.
At the very beginning
  of the plateau stage, equation~(\ref{eq:tau_ss})
  for the dense near-photospheric layers
  can be estimated as
\begin{equation}\label{eq:tau_ss_simple}
  \tau_{ss}
    \sim
    \frac{1}{\WI{u}{e}^{ss} R}
    \frac{P}{Q}
    =
    t_\mathrm{rec}
    \frac{P}{Q},
\end{equation}
  where ${t_\mathrm{rec}}$ is the classical recombination time
  \citep[p.~22]{OsterbrockFerland2006book}.
This formula is consistent with the estimates of the effective recombination time-scale
  \multicitep{UtrobinChugai2005, eq.~4; DessartHillier2007, eq.~21}.

However, a more general conclusion can be obtained from (\ref{eq:tau_ss})
  for layers of any density located both
  near and far from the photosphere.
As the constant $B$ is small
  (see Table~\ref{tab:values}, Appendix~\ref{ap:values}),
  $\tau_{ss}$ in equation~(\ref{eq:tau_ss}) can be expand as
\begin{equation}\label{eq:tau_ss_approx}
  \tau_{ss}(\tau)
    \approx
    f(\tau,A,P,s_2,R)\sqrt{\frac{\tau^{s_1}}{B}}
    +
    \mathcal{O}(\sqrt{B}).
\end{equation}
Here, $f(\tau,A,P,s_2,R)$ is a function that depends only
  on the parameters $A$, $P$, $s_2$, $R$.
We limit ourselves to calculations
  for the \emph{first} half of the photospheric phase,
  when the photoionization term $P/\tau^{s_2}$
  dominates the spontaneous emission $A\tau^{2}$,
  and an expression for $f$ can be obtained explicitly.
Then (\ref{eq:tau_ss_approx}) will be rewritten as
\begin{equation}\label{eq:tau_ss_approx2}
  \begin{multlined}
  \tau_{ss}(\tau)
    \approx
    \frac{\sqrt{{A \tau^3}/{R}}}{2(Q + A \tau^2)}
    \sqrt{\frac{P}{\tau^{{s_2}}}}
    \sqrt{\frac{\tau^{s_1}}{B}} =
    \\
    {=}
    \frac{\sqrt{{A \tau^3}/(R \tilde{B})}}{2(Q + A \tau^2)}
      \sqrt{\frac{P_{2c}(t_0)}{J_c(\nu_{L\alpha}, t_0)}}\left(\frac{t}{t_0}\right)^{\frac{s_1-s_2}{2}},
  \end{multlined}
\end{equation}
  where we used equations
  (\ref{eq:abr}), (\ref{eq:jc}), (\ref{eq:p2c2}), (\ref{eq:abr2}).
Let us rewrite this expression for the relaxation time
  using physical dimensional values:
\begin{equation}\label{eq:t_ss_approx}
  t_{ss}(t)
    \approx
    \tilde{f}(t,N_0,\WI{T}{e})
    \sqrt{\frac{P_{2c}(t_0)}{J_c(\nu_{L\alpha}, t_0)}}\left(\frac{t}{t_0}\right)^{\frac{s_1-s_2}{2}}.
\end{equation}
The minimum and maximum of the function $\tilde{f}$
  for the domain of $N_0$, $\WI{T}{e}$ (see Appendix~\ref{ap:values})
  at any given time $t \geqslant t_0$
  differ by less than an order of magnitude.
It can be concluded that the key factor that affects
  the value of the relaxation time for the whole system
  is the value and the rate of change of the external hard continuous radiation
  between the Lyman and Balmer ionization thresholds.
Because usually ${s_1 > s_2}$, the rate of change
  of the concentration of L$\alpha$ photons coming from outside
  plays the most important role for the existence of the time-dependence effect.
If there are a lot of hard photons,
  and their concentration decreases very slowly,
  then there will be no time-dependence effect.
Similar consideration can be made for the \emph{second} half
  of the photospheric phase,
  and the conclusion will be the same.

%

\section{Full kinetic system}
  \label{sec:full_system}

The equilibrium criterion can be expanded.
The reciprocal of the smallest by modulus of the Jacobi matrix eigenvalue
  can be calculated for a complete kinetic system
  that takes into account all kinds of processes.

Let us fix the external field of continuous radiation
  and find the time evolution of relaxation times
  for different kinetic systems.

A description of the modelling for various kinetic systems is given in \citet{PotashovBlinnikovUtrobin2017a}.
Calculations are carried out using the original code \code{LEVELS}
  and based on the model \texttt{R450\_M15\_Ni004\_E7} for SN~1999em
  \citep{BaklanovBlinnikovPavlyuk2005, PotashovBlinnikovUtrobin2017a}.
The physical parameters for the calculation are taken from
  the Lagrangian layer with the velocity
  $v \approx 5.8 \cdot 10^3 \,\textnormal{km} \; \textnormal{s}^{-1}$.
This layer is close to the photosphere
  at the beginning of the plateau epoch
  and its characteristics were used in the
  Table~\ref{tab:values}, column~2.
Eigendecomposition of the system Jacobi matrix is done
  by the FEAST eigensolver package
  \citep{Polizzi2009, TangPolizzi2014, KestynPolizziTang2016}.

\begin{table}
  \centering
  \caption{
  Different kinetic systems.
  }
  \label{tab:systems}
  \begin{tabular}{cccc}
    \toprule
    Model &
    Elements &
    Levels &
    \multicolumn{1}{c}{\begin{tabular}{@{}c@{}}Collisional \\ processes\end{tabular}}
    \\
    \midrule
    $M_1$ & H              &  2 & no \\
    $M_2$ & H              &  2 & yes \\
    $M_3$ & H              & 10 & yes \\
    $M_4$ & H ${+}$  He ${+}$ metals & 10 & yes \\
    \bottomrule
  \end{tabular}
\end{table}

We considered four systems presented in Table~\ref{tab:systems}.
System $M_1$ is the simplest one.
It contains pure hydrogen with
  ``two levels plus continuum'' and does
  not include any collisional processes.
This system has been considered analytically in previous sections.
System $M_2$ is similar to $M_1$,
  but with collisional processes included.
System  $M_3$ is also a pure hydrogen one,
  with collisional processes,
  but the model of the hydrogen atom contains 10 levels.
System $M_4$ is similar to $M_3$, but with metal admixtures.
For model $M_3$,
  we use a modified Sobolev approximation
  that takes into account the absorption in the continuum in the region of a line
  \citep{HummerRybicki1985, Chugai1987, Grachev1988}.
For model $M_4$,
the selective absorption of L$\alpha$ photons
  in metal lines (for example, Fe~II)
  \citep{Chugai1988b, Chugai1998}
  and overlapping lines
  \citep{Olson1982, Andronova1990, BartunovMozgovoi1987}
  is taken into account.
The metal abundances are taken from the
  \texttt{R450\_M15\_Ni004\_E7} model.
Moreover, for model $M_4$ we do not assume
  a detailed balance of the ground level with the continuum,
  and calculate the diffusive average intensity
  of the Lyman continuum
  in a self-consistent manner
  \citep[eq.~5]{UtrobinChugai2005}.

Fig.~\ref{fig:tss_full} shows the results of the calculation.
The thin solid line ``ph'' corresponds
  to the analytical estimation
  obtained with the formula~(\ref{eq:tau_ss}).
This is the same line as that one designated by ``ph'' in Fig.~\ref{fig:tss}.
Numerical calculation of the model $M_1$
  (thick dashed line) agrees well with it.
The thick solid curve $M_2$ deviates from $M_1$ during the first 10~d after $t_0$.
Collisional processes
  decrease the relaxation time
  only at the very beginning of the photospheric phase.
Additional levels of $M_3$ (dotted line) slightly decrease the relaxation time.
Taking metals into account in $M_4$ (bold dashed line)
  leads to a significant decrease of the relaxation time
  and hence to a weakening of the time-dependence effect.
Because the relaxation time is still much longer than the duration of the photospheric phase,
  this is not enough to completely cancel the non-stationarity.
This result was already obtained earlier in \cite{PotashovBlinnikovUtrobin2017a},
  where it was shown that the increase of the metal concentration
  in the envelope leads to a weakening of the time-dependence effect in kinetics
  when the continuous radiation is \emph{fixed}.
This allows us to say that the metal admixtures
  play the most important role in the non-stationarity
  among other effects studied in this section.

Summarizing the statements above
  with the findings of Section~\ref{sec:frozen_system},
  we can conclude the following.
The concentration of L$\alpha$ photons that come from outside,
  as well as the change of this concentration
  and the rate of the selective absorption of L$\alpha$
  photons in metal lines
  are the most significant factors affecting the non-stationarity effect.
The greater the number of external L$\alpha$ photons
  and the faster their absorption by metals,
  the weaker the effect of time-dependent ionization.

The metallicity of the envelope
  strongly affects the flux in the U band,
  and this complicates the situation.
The emergent UV flux is significantly enhanced
  at low metallicity, owing to the reduction of line blanketing
  \cite[Goldstein \& Blinnikov in preparation;][]{DessartHillierWaldmanEtal2013}.
On one hand,
  this means that the low metallicity
  leads to an increase of the number
  of external L$\alpha$ photons
  and, consequently, decreases the relaxation time of the system.
On the other hand, there is an opposite effect of the relaxation time increase
  owing to the reduced number of absorbers.
Therefore, the final answer to the question of
  whether the time-dependent ionization effect is important
  requires checking the equilibrium criterion described here.
At least for SN~1999em, our calculation shows
  the importance of accounting for
  the time-dependence in kinetics
  (see Fig.~\ref{fig:tss_full}, model $M_4$).

\section{Discussion and Conclusions}
  \label{sec:discussions_conclusions}

There are two characteristic time-scales in the SN ejecta.
One defines the time of relaxation to equilibrium conditions.
This time equals the reciprocal of the
  smallest by modulus of Jacobi matrix eigenvalue
  for a complete kinetic system that takes into account
  all kinds of processes at any time
  (not only during the photospheric phase).
The second one is the typical time-scale
  for the change of SN envelope parameters.

We come to the key conclusion of the paper.
If the ratio of the two time-scales is small the system is in equilibrium.
In this case one can investigate the steady-state algebraic approximation
  instead of the real time-dependent system.
If this criterion of equilibrium is not satisfied,
  then a time-dependent effect is important,
  and the real atomic level populations
  deviate from the equilibrium values.
In this case,
  the replacement of the original time-dependent system
  with the algebraic steady-state one is incorrect.

\section*{Acknowledgements}

The authors are grateful to S.I.~Blinnikov,
  V.P.~Utrobin and P.V.~Baklanov for useful discussions.
The authors are also thankful to E.I.~So\-ro\-ki\-na for
  thorough reading the manuscript before publishing.
M.Sh.~Po\-ta\-shov thanks S.G.~Moiseenko and G.S.~Bisnovaty-Kogan
  for repeated invitations to the school-seminar organized by them in the town of Tarusa,
  where the results of this paper were presented.
We are grateful to the anonymous referee for valuable comments.
The work of MP
  was partially supported by the grant of
  the Russian Foundation for Basic Research~(RFBR), grant~19--02--00567;
  the numerical calculations in the Section~\ref{sec:full_system}
  were funded by the Russian Science Foundation, grant~19--12--00229.
The work of AY was supported by the RFBR, grant~18--29--21019.



\appendix

\section{Detailed balance of the ground level population of hydrogen}
  \label{ap:balance}

Let us consider the photo-processes that influence
  the electronic population of the ground level of the hydrogen atom.
We assume that there are admixtures of other elements in the envelope.
Let us introduce two assumptions.
First, the main contribution to the opacity
  in the frequency band of the Lyman continuum
  ${\nu \geqslant \nu_{LyC}}$ is
  due to the free-bound processes and caused mainly by neutral hydrogen.
Secondly, the optical depth in this band is very large.

Let us write down the coefficients of emission and true absorption,
  corrected for induced emission using the Einstein-Milne relations for continua
  \citep{Mihalas1978book, HubenyMihalas2014book}:
\begin{equation}\label{ap:eq:chi}
  \chi_{\nu} = (n_1 - n_1^{*} e^{-\tfrac{h \nu}{k \WI{T}{e}}})\,\alpha_{1c}(\nu),
\end{equation}
\begin{equation}\label{ap:eq:eta}
  \eta_{\nu} = n_1^{*} (1 - e^{-\tfrac{h \nu}{k \WI{T}{e}}})\,\alpha_{1c}(\nu)B_{\nu}(\WI{T}{e}).
\end{equation}
Here $\alpha_{1c}(\nu)$ is the photoionization cross section from the first level,
  $B_{\nu}(\WI{T}{e})$ is the intensity
  of blackbody radiation for the electron temperature $\WI{T}{e}$, and
\begin{equation}\label{ap:eq:n1s}
  n_1^{*} = \WI{n}{e} \WI{n}{p} \Phi_{\mathrm{Saha}}(\WI{T}{e})
\end{equation}
  is the local thermodynamic equilibrium (LTE) value of $n_1$
  computed with the Saha equation using the \emph{actual} values of
  the electron and proton number densities $\WI{n}{e}$ and $\WI{n}{p}$.

For an optically thick medium,
  the transport equation in the continuum is solved as
  $J_c(\nu, t) = S_c(\nu, t)$,
  where $S_c(\nu, t)$ is a source function,
  defined as
\begin{equation}\label{ap:eq:sc}
  S_c(\nu, t) = \frac{\eta_{\nu}}{\chi_{\nu}}.
\end{equation}
Substituting (\ref{ap:eq:chi}) and (\ref{ap:eq:eta}) into (\ref{ap:eq:sc})
  we have
\begin{equation}\label{ap:eq:jc}
  J_c(\nu, t) = \frac{(1 - e^{-\tfrac{h \nu}{k \WI{T}{e}}})}{\frac{n_1}{n_1^{*}} - e^{-\tfrac{h \nu}{k \WI{T}{e}}}}B_{\nu}(\WI{T}{e}).
\end{equation}

In accordance with
  \citet{Mihalas1978book}
  and
  \citet[p.~273]{HubenyMihalas2014book},
  the photoionization rate from the ground level is the integral
\begin{equation}\label{ap:eq:p1c}
  n_1 P_{1c}(t) = 4\pi n_1\mkern-10mu\int\limits_{\nu_{LyC}}^{\infty}\mkern-10mu\frac{\alpha_{1c}(\nu)}{h\nu} J_c(\nu, t) d\nu.
\end{equation}
The photorecombination rate looks like
\begin{equation}\label{ap:eq:rc1}
  \WI{n}{e} \WI{n}{p} R_{c1}(t)
    = 4\pi \WI{n}{e} \WI{n}{p}\,\Phi_{\mathrm{Saha}}(\WI{T}{e})
    \times
    \mkern-10mu\int\limits_{\nu_{LyC}}^{\infty}\mkern-10mu\frac{\alpha_{1c}(\nu)}{h\nu}\Big(\frac{2h\nu^3}{c^2}
    +
    J_c(\nu, t)\Big)e^{-\frac{h\nu}{k\WI{T}{e}}}d\nu.
\end{equation}
Substituting equation (\ref{ap:eq:jc}) into (\ref{ap:eq:p1c}) and (\ref{ap:eq:rc1}),
  and taking into account (\ref{ap:eq:n1s}) gives $n_1 P_{1c}(t) = \WI{n}{e} \WI{n}{p} R_{c1}(t)$.
Thus, in the general case, with admixtures,
  the photoionization rate from the ground level of hydrogen
  and the radiative recombination rate to the ground level completely coincide
  and the ground level of hydrogen is in detailed balance with the continuum.

\section{Typical values}
  \label{ap:values}

Let us write down the typical
  values of the physical parameters
  of the SN envelope and its radiation,
  based on calculations for SN~1999em
  \citep{BaklanovBlinnikovPavlyuk2005, PotashovBlinnikovUtrobin2017a},
  which is a standard SN~IIP.
The computations were performed with the
  radiation-hydrodynamical code \code{STELLA}.
We will mark values that correspond
  to the near-photospheric layers with ``$\mathrm{ph}$''
  and to the outer, high-velocity layers with ``$\mathrm{out}$''.

The initial time is ${t_0 \approx 20}$~d after the explosion
  in the core.

The two-photon ${2\mathrm{s} \to 1\mathrm{s}}$ decay rate is taken from
  \citet{NussbaumerSchmutz1984}.
We consider two $l$-sublevels ${2\mathrm{s}}$ and ${2\,\mathrm{p}}$
  as a single super-level 2
  \citep{HubenyLanz1995}
  and assume that the populations of the sublevels are proportional to their
  statistical weights ${g_{2\mathrm{s}}}$, ${g_{2\mathrm{p}}}$.
This leads to
\begin{equation*}
  Q \approx 8.2249 \frac{g_{2\mathrm{s}}}{g_{2\mathrm{s}}+g_{2\mathrm{p}}}\,t_0 \approx
  2.74\,t_0
  \approx 3.5 \cdot 10^6.
\end{equation*}

The initial number densities of  gas are
  ${N_0(\mathrm{ph}) \approx 10^{11} \,\mathrm{cm}^{-3}}$
  and
  ${N_0(\mathrm{out}) \approx 10^8 \,\mathrm{cm}^{-3}}$.
It follows that for the constant $A$
  defined in (\ref{eq:abr}, \ref{eq:abr2}) we have corresponding values
\begin{equation*}
  A(\mathrm{ph}) \approx 3\cdot10^4,\quad
  A(\mathrm{out}) \approx 3\cdot10^7.
\end{equation*}
Typical values of the intensity of the continuum radiation
  at the initial moment
  at the L$\alpha$ frequency
  is taken from the calculation of \code{STELLA}:
\begin{gather*}
  J_c(\nu_{L\alpha}, t_0, \mathrm{ph}) \approx 10^{-8} \,\mathrm{\frac{erg}{cm^2\,s\,Hz}}.
  \\
  J_c(\nu_{L\alpha}, t_0, \mathrm{out}) \approx 10^{-12}\,\mathrm{\frac{erg}{cm^2\,s\,Hz}},
\end{gather*}
Thus, for $B$ defined in (\ref{eq:abr}, \ref{eq:abr2}) we have
\begin{equation*}
  B(\mathrm{ph}) \approx 6\cdot10^{-3},\quad
  B(\mathrm{out}) \approx 6\cdot10^{-4}.
\end{equation*}

The photoionization coefficient from the second level $P_{2c}$
  at the initial moment is also
  taken from \code{STELLA} calculations:
\begin{equation*}
  P_{2c}(t_0,\mathrm{ph}) \approx 10^{5} \,\mathrm{s}^{-1},\quad
  P_{2c}(t_0,\mathrm{out}) \approx 4\cdot10^{3} \,\mathrm{s}^{-1}.
\end{equation*}
Then we get from (\ref{eq:abr2}):
\begin{equation*}
  P(\mathrm{ph}) \approx 2\cdot10^{11},\quad
  P(\mathrm{out}) \approx 7\cdot10^{9}.
\end{equation*}

The effective photorecombination coefficient to the second level
  at the initial moment can be estimated from
  \citet[Table 1]{Hummer1994}
  using the material temperatures typical for a supernova envelope
  ${\WI{T}{e} \approx 3000 - 7000\,\mathrm{K}}$ as
  ${R_{2c} \approx 3 \cdot 10^{-13} \,\mathrm{\frac{cm^3}{s}}}$.
Then we will have for $R$ from (\ref{eq:abr}, \ref{eq:abr2}):
\begin{equation*}
  R(\mathrm{ph}) \approx 5\cdot10^{4},\quad
  R(\mathrm{out}) \approx 50.
\end{equation*}
Values for the power exponents taken from \code{STELLA} calculations are
\begin{gather*}
  s_1(\mathrm{ph}) \approx 24,\quad
  s_1(\mathrm{out}) \approx 21,
  \\
  s_2(\mathrm{ph}) \approx 6,\quad
  s_2(\mathrm{out}) \approx 4.
\end{gather*}
Because in the optically thin atmosphere
  ${s_1 = 2}$ and ${s_2 = 2}$,
  in general case, the domains of these exponents are
  ${s_1 \geqslant 2}$ and ${s_2 \geqslant 2}$.

\section{The tube}In general case, we limit the domain of these exponents to
  ${s_1 \geqslant 2}$ and ${s_2 \geqslant 2}$.

  \label{ap:tube}

Let us prove the dissipativity of the system
  (\ref{eq:kinetics:td2:u1}, \ref{eq:kinetics:td2:ue})
  in another way.
Suppose that we know some particular bound solution
  $0 {<} \tilde{u}_1 {\leqslant} 1$ and $0 {\leqslant} \WI{\tilde{u}}{e} {\leqslant} 1$
  (\emph{unperturbed motion})
  of a nonautonomous nonlinear differential system
  (\ref{eq:kinetics:td2:u1}, \ref{eq:kinetics:td2:ue}).
The change of variables
  ${x_1 = u_1 - \tilde{u}_1}$ and ${\WI{x}{e} = \WI{u}{e} - \WI{\tilde{u}}{e}}$
  transforms the system into the form of \emph{perturbed motion} system
  \multicitep{Demidovich1967book, p.~234; Khalil2002book, p.~147}:
\begin{equation}\label{eq:kinetics:x1}
  \dot{x}_1 = -(x_1+\WI{x}{e})\;Q
    - \frac{(1-\WI{\tilde{u}}{e})x_1+\tilde{u}_1 \WI{x}{e}}{\tilde{u}_1(\tilde{u}_1+x_1)}A\tau^2,
\end{equation}
\begin{equation}\label{eq:kinetics:xe}
  \WI{\dot{x}}{e} = - (x_1+\WI{x}{e})\frac{P}{\tau^{s_2}}
    -\WI{x}{e}\;(2\WI{\tilde{u}}{e}+\WI{x}{e})\frac{R}{\tau^3}.
\end{equation}
The trivial solution ${x_1 = 0}$, ${\WI{x}{e} = 0}$
  is an equilibrium point of the transformed system for any $\tau \geqslant 1$.

Consider the Lyapunov function candidate
\begin{equation}\label{eq:v}
  V(\tau,x_1,\WI{x}{e}) =
    x_1^2 \frac{(1-{\WI{\tilde{u}}{e}})}{{\tilde{u}_1}}
    + 2 x_1 \WI{x}{e}
    + \WI{x}{e}^2 \left(1+d\right).
\end{equation}
The function $V$ is positively defined at $d > 0$
  and it is an elliptic paraboloid.
The derivative of $V$ along the trajectories of the system
  (\ref{eq:kinetics:x1}, \ref{eq:kinetics:xe}) is given by
\begin{equation*}
  \dot{V}(\tau, x_1, \WI{x}{e}) = \frac{\partial V}{\partial \tau}
    + \frac{\partial V}{\partial x_1} \dot{x}_1 + \frac{\partial V}{\partial \WI{x}{e}} \WI{\dot{x}}{e},
\end{equation*}
  where $\dot{x}_1$ and $\WI{\dot{x}}{e}$
    are (\ref{eq:kinetics:x1}) and (\ref{eq:kinetics:xe}), respectively.
At large time and at \emph{small} $x_1$ and $\WI{x}{e}$ it expresses as
\begin{equation}\label{eq:dv}
  \dot{V}(\tau, x_1, \WI{x}{e}) =
    -\frac{2 A\;\tau^2}{{\tilde{u}_1}}
    \Bigg[\frac{(1-{\WI{\tilde{u}}{e}})(1-{\WI{\tilde{u}}{e}}-{\tilde{u}_1})\;x_1^2}{2\;{\tilde{u}_1}^2}
    +
    \left(\frac{(1-{\WI{\tilde{u}}{e}})}{{\tilde{u}_1}}x_1+\WI{x}{e}\right)^2\Bigg] + \mathcal{O}(\tau).
\end{equation}
This function is negative semidefinite in the domain
  ${\Omega = (0 < \tilde{u}_1 {+} x_1 \leqslant 1, 0 < \WI{\tilde{u}}{e
  } {+} \WI{x}{e} \leqslant 1)}$.

From the Lyapunov direct method and (\ref{eq:v}, \ref{eq:dv})
  it follows that the trivial solution of system (\ref{eq:kinetics:x1}, \ref{eq:kinetics:xe})
  is stable for small perturbations of the initial conditions
  for any positive $d$.
The stability problem for this system
  can be solved in this way on the half-axis
  ${\tau > \tau_1}$
  with $\tau_1 \geqslant 1$.
The stability on a half-axis ${\tau > 1}$ is obtained taking into account
  the theorem of continuous dependence on the parameter
  \citep[p.~95]{Khalil2002book}
  for the solution on a finite interval ${1 \leqslant \tau \leqslant \tau_1}$.

The analysis of the function (\ref{eq:v}) reveals that the system
  (\ref{eq:kinetics:x1}, \ref{eq:kinetics:xe})
  is  dissipative.
Indeed, Section~\ref{sec:large_time} shows
  that ${\tilde{u}_1 \approx 1-\WI{\tilde{u}}{e}}$
  at large time.
Then (\ref{eq:dv}) can be rewritten as
\begin{equation*}
  \dot{V}(\tau, x_1, \WI{x}{e}) = -\frac{2 A\;\tau^2}{\tilde{u}_1}(x_1+\WI{x}{e})^2 + \mathcal{O}(\tau).
\end{equation*}
This function is a parabolic cylinder
  with the zero values along the line $l$
  defined by the equation ${x_1 = -\WI{x}{e}}$.
This means that one can specify a value $\mu$
  so that the set $l$ does not belong
  to the $\tilde{\Omega}$ domain
  defined as the difference between the $\Omega$ domain
  and the domain specified by the inequality
\begin{equation*}
  \sup_{\tau \geqslant 1} V(\tau,x_1,\WI{x}{e}) \leqslant \mu.
\end{equation*}
  for all moments of time ${\tau \geqslant 1}$.
Thus, the function $\dot{V}(\tau, x_1, \WI{x}{e})$
  is negative definite in $\tilde{\Omega}$.
Therefore, according to the theorem of Yoshizawa
  \multicitep{Demidovich1967book, p.~290; KuncevichLychak1977book, p.~47-48}, the
  perturbed motion of the system is dissipative.
Hence, any solution of the equation,
  starting from arbitrary time,
  eventually will lie inside a cylinder of a non-zero radius
  and will never go out of it.

The given proof of dissipativity assumes
  small perturbations $x_1$ and $\WI{x}{e}$.
Let us prove by contradiction that dissipativity is true for any
  $x_1$, $\WI{x}{e}$ in $\Omega$.
Note additionally that the solutions of the initial value problem
  (\ref{eq:kinetics:td2:u1}, \ref{eq:kinetics:td2:ue}, \ref{eq:kinetics:td2:init})
  are bounded.
If one suggests the presence of two cylinders
  then one inevitably obtains two close-lying bounded solutions
  that belong to different cylinders.
Contra this, we showed earlier that all close-lying solutions
  must lie in one cylinder owing to dissipativity.
This contradiction yields the proof of uniqueness of the cylinder-tube.

\begin{figure}
  \centering
  \begin{minipage}{0.5\textwidth}
    \includegraphics[width=\textwidth]{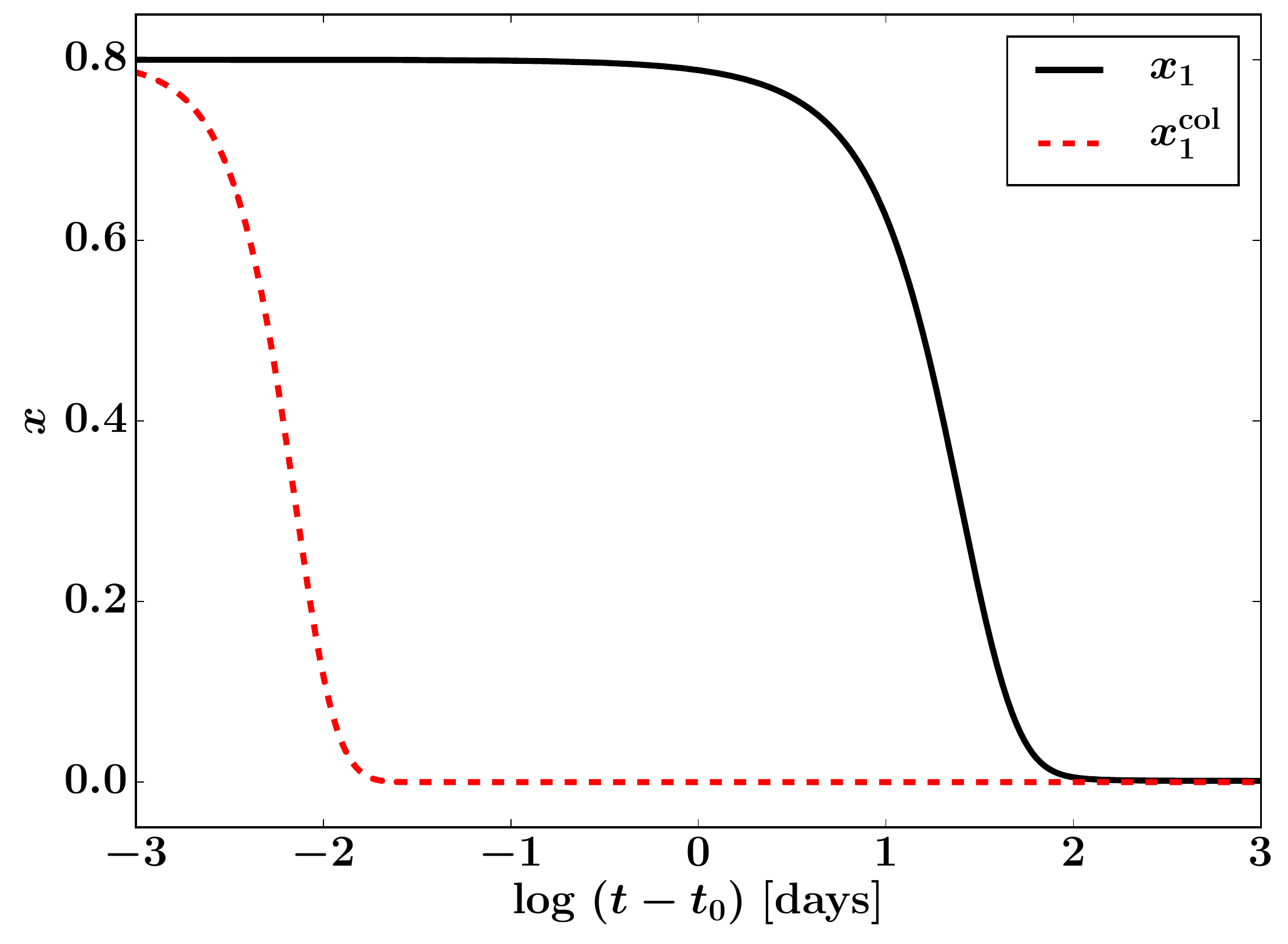}
  \end{minipage}
  \caption {
    The functions $x_1$ and $x_1^{\mathrm{col}}$
      depending on physical time with ${t_0 {=} 20}$~d.
      $x_1^{\mathrm{col}}$ is the solution of the perturbed motion system
      (\ref{eq:kinetics:x1}, \ref{eq:kinetics:xe})
      with collisional processes taken into account.
    The calculation has been performed
      for an optically thick medium
      and physical parameters typical for
      the near-photospheric layers
      (see Table~\ref{tab:values} column~2, Appendix~\ref{ap:values}).
  }
  \label{fig:x1_col}
\end{figure}

The theorem of Yoshizawa does not give practically useful estimates
  of the convergence time of the solutions of a dissipative system
  into a semi-infinite cylinder-tube.
By restricting our analysis to the
  numerical estimates of this time,
  we can solve the system
  (\ref{eq:kinetics:x1}, \ref{eq:kinetics:xe}) with the initial conditions
  ${x_1(1) \approx \tilde{u}_1(1)}$,
  ${\WI{x}{e}(1) = 1 - \WI{\tilde{u}}{e}(1)}$.
It was shown in Section~\ref{sec:large_time}
  that the initial condition of $u_2$ does not affect the solution.
Because ${u_2 = 1- u_1 - \WI{u}{e}}$,
  then from (\ref{eq:kinetics:td3:u2:sol}) it follows that the
  expression ${-(x_1+\WI{x}{e})}$
  decreases exponentially as $e^{-G_3(\tau)}$.
Therefore, without a loss of generality,
  only the behaviour of $x_1$ is considered below.
Fig.~\ref{fig:x1_col} shows that the convergence time
  of the solutions of the dissipative system to the cylinder-tube is large
  for an optically thick atmosphere and physical parameters typical for
  the near-photospheric layers of SN~IIP
  and equal to about ${\sim 10}$~d.
The situation changes dramatically
  if one takes into account collisional processes.
For example, in Fig.~\ref{fig:x1_col} we present
  the results of calculation where
  collisional processes have been taken into
  account under the same conditions.
The convergence time in this case is only thousands of seconds.
If one neglects the width of the cylinder-tube
  (${x_1^{\mathrm{col}} \ll 1}$ at time greater than 1~d)
  one can say that the system ``forgets'' the initial conditions!

%


\end{document}